\documentclass[useAMS,usenatbib]{mn2e}
\usepackage{longtable}
\usepackage{graphicx}
\usepackage{lscape}
\usepackage{rotating}
\newcommand{\mnras}{MNRAS}
\newcommand{\aj}{AJ}
\newcommand{\apj}{ApJ}
\newcommand{\apjs}{ApJ}
\newcommand{\apjl}{ApJ}
\newcommand{\aap}{A\&A}

\newcommand{\pasp}{PASP}

\newcommand{\memsai}{Memorie della Societa Astronomica Italiana}

\newcommand{\Mwd}{\mbox{$M_\mathrm{WD}$}}

\newcommand{\Twd}{\mbox{$T_\mathrm{WD}$}}

\newcommand{\Msun}{\mbox{$\mathrm{M}_{\odot}$}}

\newcommand{\Teff}{\mbox{$T_{\mathrm{eff}}$}}

\newcommand{\Lines}[3]{\Ion{#1}{#2}\,$\lambda\lambda$\,#3}
\newcommand{\Ion}[2]{#1{\,\scriptsize #2}}

\newcommand{\kms}{\mbox{$\mathrm{km\,s^{-1}}$}}

\title[The  SDSS  WDMS  binary   catalogue]  {The  SDSS  spectroscopic
  catalogue of white dwarf-main sequence binaries: new identifications
  from DR\,9--12}

\author[A.   Rebassa-Mansergas  et  al.]{A.   Rebassa-Mansergas$^{1}$,
  J.J.  Ren$^{2}$,  S.G.  Parsons$^{3}$, B.T.   G\"ansicke$^{4}$, M.R.
  Schreiber$^{3,5}$,   \newauthor  E.   Garc\'ia-Berro$^{1,6}$,  X.-W.
  Liu$^{2,7}$, D. Koester$^{8}$\\
$^{1}$ Departament de F\'\i sica, Universitat Polit\`ecnica de
  Catalunya, c/Esteve Terrades 5, 08860 Castelldefels, Spain\\
$^{2}$ Department of Astronomy, Peking University, Beijing 100871,
 P.\,R.\,China\\
$^{3}$ Instituto de F\'\i sica y Astronom\'\i a, Universidad de
 Valpara\'\i so, Avenida Gran Breta\~na 1111, Valpara\'\i so, Chile \\
$^{4}$ Department of Physics, University of Warwick, Coventry CV4 7AL,
 UK \\
$^{5}$ Millenium Nucleus "Protoplanetary Disks in ALMA Early Science",
 Universidad de Valpara\'\i so, Avenida Gran Breta\~na 1111,
 Valpara\'\i so, Chile\\
$^{6}$ Institute for Space Studies of Catalonia, c/Gran Capit\`a 2--4,
  Edif. Nexus 201, 08034 Barcelona, Spain\\
$^{7}$  Kavli   Institute  for  Astronomy  and   Astrophysics,  Peking
  University, Beijing 100871, P.\,R.\,China\\
$^{8}$ Institut f\"ur Theoretische Physik und Astrophysik, University
 of Kiel, 24098 Kiel, Germany
}

\begin{document}
\date{Accepted 2015. Received 2015; in original form 2015}
\pagerange{\pageref{firstpage}--\pageref{lastpage}} \pubyear{2015}
\maketitle

\begin{abstract}
We present an updated version  of the spectroscopic catalogue of white
dwarf-main sequence (WDMS) binaries from  the Sloan Digital Sky Survey
(SDSS).  We identify  939 WDMS binaries within the  data releases (DR)
9--12 of  SDSS plus  40 objects  from DR\,1--8 that  we missed  in our
previous  works,  646   of  which  are  new.   The   total  number  of
spectroscopic SDSS WDMS binaries increases to 3294. This is by far the
largest  and most  homogeneous  sample of  compact binaries  currently
available.   We  use a  decomposition/fitting  routine  to derive  the
stellar  parameters  of  all  systems  identified  here  (white  dwarf
effective temperatures,  surface gravities  and masses,  and secondary
star  spectral  types).  The  analysis  of  the corresponding  stellar
parameter distributions shows that the  SDSS WDMS binary population is
seriously  affected  by  selection   effects.   We  also  measure  the
\Lines{Na}{I}{8183.27,  8194.81}  absorption   doublet  and  H$\alpha$
emission  radial velocities  (RV) from  all SDSS  WDMS binary  spectra
identified  in  this  work.   98  objects  are  found  to  display  RV
variations, 62 of which are new.  The RV data are sufficient enough to
estimate the orbital periods of three close binaries.

\end{abstract}


\label{firstpage}

\section{Introduction}
\label{s-intro}

A large  fraction of main sequence  stars are found in  binary systems
\citep{duquennoy+mayor91-1,  raghavanetal10-1,  yuanetal15-1}.  It  is
expected  that $\sim$25  per cent  of all  main sequence  binaries are
close enough to begin mass transfer interactions when the more massive
star   becomes   a   red   giant   or   an   asymptotic   giant   star
\citep{willems+kolb04-1}.   This  has  been  observationally  verified
e.g. by  \citet{farihietal10-1} and by  \citet{nebotetal11-1}. Because
the  mass transfer  rate generally  exceeds the  Eddington limit,  the
secondary star is not able to accrete the transferred material and the
system evolves through a common envelope phase. That is,  the core of
the  giant and  the  main  sequence companion  orbit  within a  common
envelope formed  by the outer layers  of the giant star.   Drag forces
between the  two stars and the  envelope lead to the  shrinkage of the
orbit and  therefore to  the release of  orbital energy.   The orbital
energy is deposited into the envelope  and is eventually used to eject
it \citep{webbink08-1}.   The outcome of common  envelope evolution is
hence a close binary formed by the core of the giant star (which later
becomes a  white dwarf) and a  main sequence companion, i.e.   a close
white dwarf-main  sequence (WDMS) binary. These  are commonly referred
to as  post-common envelope  binaries (PCEBs). The  remaining $\sim$75
per  cent of  main sequence  binaries are  wide enough  to avoid  mass
transfer interactions.  In  these cases the more  massive stars evolve
like  single stars  until they  eventually become  white dwarfs.   The
orbital separations of such WDMS binaries  are similar to those of the
main sequence binaries from which they descend.

During  the last  years we  have mined  the Sloan  Digital Sky  Survey
\citep[SDSS;][]{yorketal00-1} spectroscopic data base  to build up the
largest,  most homogeneous  and  complete catalogue  of WDMS  binaries
\citep{rebassa-mansergasetal10-1,           rebassa-mansergasetal12-1,
  rebassa-mansergasetal13-2}, which contains 2316  systems as for data
release (DR) 8.   Observational studies led by our  team have resulted
in the  identification of 1050 wide  SDSS WDMS binaries and  206 PCEBs
\citep{rebassa-mansergasetal07-1,                   schreiberetal08-1,
  schreiberetal10-1,  rebassa-mansergasetal11-1},  for which  we  have
measured  the orbital  period of  90 \citep{rebassa-mansergasetal08-1,
  nebotetal11-1, rebassa-mansergasetal12-2}.

The analysis  of the entire  SDSS WDMS binary  sample, as well  as the
sub-samples of close PCEBs and widely separated WDMS binaries have led
to numerous advances in several  fields of modern astrophysics.  These
include for  example: constraining  current theories of  close compact
binary     evolution      \citep{zorotovicetal10-1,     davisetal10-1,
  zorotovicetal11-2,  rebassa-mansergasetal12-2};  demonstrating in  a
robust way  that the majority of  low-mass white dwarfs are  formed in
binaries  \citep{rebassa-mansergasetal11-1}; constraining  the pairing
properties  of  main  sequence stars  \citep{ferrario12-1};  providing
robust   observational  evidence   for   disrupted  magnetic   braking
\citep{schreiberetal10-1};   constraining  the   rotation-age-activity
relation       of       low-mass       main       sequence       stars
\citep{rebassa-mansergasetal13-1}; studying the statistical properties
of    the   PCEB    population    using    Monte   Carlo    techniques
\citep{toonen+nelemans13-1,    camachoetal14-1,    zorotovicetal14-2};
analysing  why  the  average  mass  of  white  dwarfs  in  cataclysmic
variables  is significantly  larger than  the average  mass of  single
white     dwarfs      \citep{zorotovicetal11-1,     schreiberetal16-1,
  nelemansetal15-1}; testing hierarchical probabilistic models used to
infer  properties  of  unseen  companions  to  low-mass  white  dwarfs
\citep{andrewsetal14-1}; detecting new gravitational wave verification
sources  \citep{kilicetal14-1};  analysing   the  interior  structural
effects  of  common envelope  evolution  on  low-mass pulsating  white
dwarfs \citep{hermesetal15-1}.  In addition, many eclipsing SDSS PCEBs
have    been    identified    \citep{nebotetal09-1,    pyrzasetal09-1,
  pyrzasetal12-1,  parsonsetal13-1, parsonsetal15-1}  which are  being
used to  test theoretical mass-radius  relations of both  white dwarfs
and    low-mass    main   sequence    stars    \citep{parsonsetal12-1,
  parsonsetal12-2}, as  well as the existence  of circumbinary planets
\citep{zorotovic+schreiber13-1, parsonsetal14-1, marshetal14-1}.

The motivation of this paper  is to update our spectroscopic catalogue
of SDSS WDMS binaries by  searching for new identifications within the
third        survey        generation        of        the        SDSS
\citep[SDSS\,III,][]{eisensteinetal11-1} to  thus increase  the number
of  systems available  for follow-up  studies\footnote{The SDSS  DR\,8
  contains the first  set of spectra observed by  SDSS\,III, which are
  not considered here because we have already identified WDMS binaries
  within this DR.   Thus, in this paper we will  consider SDSS\,III as
  all  data collected  from DR\,9  to  DR\,12.}.

\section{The SDSS\,III survey}

In this work we search for WDMS binaries within the spectroscopic data
base  of the  SDSS\,III,  i.e.  the  ninth \citep{ahnetal12-1},  tenth
\citep{ahnetal14-1}, eleventh and  twelfth \citep{alametal15-1} DRs of
the SDSS.  In particular, we  mined the optical  ($R\sim$2000) spectra
obtained  by  the  Sloan  Exploration of  Galactic  Understanding  and
Evolution 2 survey (SEGUE-2; C.  Rockosi et al.  2015, in preparation)
and      the     Baryon      Oscillation     Spectroscopic      Survey
\citep[BOSS;][]{dawsonetal13-1}.

BOSS uses  a new set of  spectrographs with 1000 fibres  available per
exposure.   The main  goal  of the  survey is  to  obtain spectra  for
galaxies and quasars selected from the  SDSS imaging data to study the
baryon  oscillation   feature  in  the  clustering   of  galaxies  and
Lyman-$\alpha$ absorption along the line  of sight to distant quasars.
BOSS uses a different target  selection criteria for targeting quasars
\citep{rossetal12-1} than the one  employed by the SDSS\,I/II surveys.
Thus, whilst  SDSS\,I/II targeted objects  with $i <$ 19.1,  BOSS went
down to  $g <$ 22.0  in order to  give a significantly  higher surface
density of targets than  in SDSS\,I/II \citep{rossetal12-1}.  BOSS has
obtained a total  number of 4,571,520 spectra.   Because WDMS binaries
and quasars overlap in colour space \citep{smolcicetal04-1}, we expect
a  large number  of  WDMS binaries  being  (most likely  accidentally)
observed   by  the   BOSS   survey.   SEGUE-2   used  the   SDSS\,I/II
spectrographs  to obtain  spectra of  stars at  high and  low Galactic
latitudes  to   study  Galactic   structure,  dynamics,   and  stellar
populations.  SEGUE-2 gathered a total of 155,520 spectra.

\section{Identification of WDMS binaries}
\label{s-ident}

\begin{figure}
\centering
\includegraphics[width=\columnwidth]{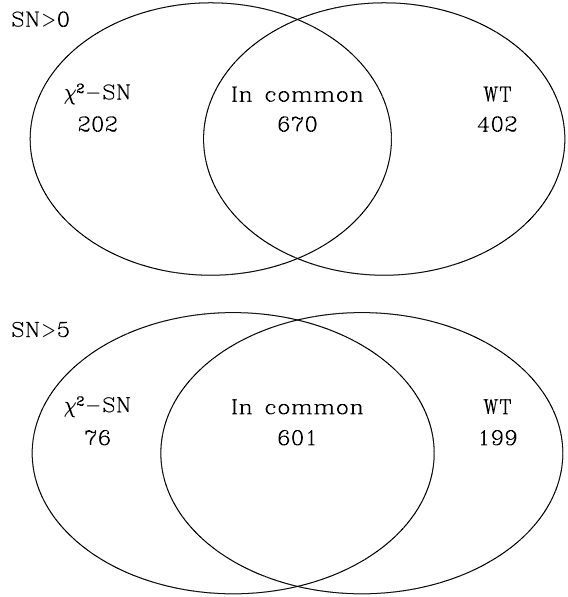}
\caption {Top panel: Venn diagram  for WDMS binaries identified by the
  $\chi^2$-SN (left)  and WT (right)  methods. Bottom panel:  the same
  but considering only spectra of SN$>$5.}
\label{f-venn}
\end{figure}

In \citet{rebassa-mansergasetal10-1}  we developed a routine  based on
reduced   $\chi^2$   and   signal-to-noise  (SN)   ratio   constraints
($\chi^2$-SN  method hereafter)  to identify  spectroscopic SDSS  WDMS
binaries.   As in  our previous  works, we  use this  method here  for
identifying WDMS binaries observed by the SDSS\,III survey.

In a first  step we $\chi^2$-fit all SDSS\,III spectra  with a grid of
163  WDMS  binary templates  covering  a  wide  range of  white  dwarf
effective   temperatures  (6000--100000\,K)   and  surface   gravities
(6.5--9.5   dex),   and   secondary  star   M-dwarf   spectral   types
(M0--M9)\footnote{Due to  selection effects the vast  majority of SDSS
  WDMS   binaries   contain   a  low-mass   M-dwarf   companion,   see
  \citealt{rebassa-mansergasetal10-1}.}   and  represent  the  reduced
$\chi^2$ values as a function of  SN ratio for each template.  We then
define an equation of the form
\begin{equation}
\chi^2_\mathrm{max}=a\times\mathrm{SN}^{b},
\end{equation}
\noindent and  consider as  WDMS binary  candidates all  objects below
this curve ($\chi^2_\mathrm{spec}<\chi^2_\mathrm{max}$), where $a$ and
$b$   are   free   parameters    defined   for   each   template   and
$\chi^2_\mathrm{spec}$  is the  $\chi^2$ that  results from  fitting a
considered     spectrum     with     a     given     template     (see
\citealt{rebassa-mansergasetal10-1} for a  complete description of the
method).  The  form of Eq.\,(1)  is defined to account  for systematic
errors  between  template  and  observed spectra,  which  become  more
important the larger the SN ratio is.  This exercise resulted in 9,593
WDMS binary candidate spectra which  were visually inspected.  We also
inspected their SDSS images for morphological problems and made use of
the  GALEX ultraviolet  \citep{martinetal05-1, morrisseyetal05-1}  and
UKIDSS infrared  \citep{dyeetal06-1, hewettetal06-1, lawrenceetal07-1}
magnitudes to  probe for the presence  of excess flux at  the blue and
red ends of the SDSS spectra,  respectively.

We found 872 of the 9,593 selected spectra to be genuine WDMS binaries
and  we considered  47 spectra  as  WDMS binary  candidates.  In  what
follows, we  flag an  object to  be a WDMS  binary candidate  when the
spectrum is that of a white dwarf (main sequence star) displaying some
red  (blue) flux  excess which  cannot be  confirmed to  arise from  a
companion due to the lack of GALEX (UKIDSS) magnitudes.

The  above outlined  $\chi^2$-SN method  has  been proven  to be  very
efficient  at  identifying  WDMS binaries,  achieving  a  completeness
(defined as the  ratio between the number of SDSS  WDMS binary spectra
we  have successfully  identified and  the total  number of  SDSS WDMS
binary      spectra     observed)      around     96      per     cent
\citep{rebassa-mansergasetal10-1,           rebassa-mansergasetal12-1,
  rebassa-mansergasetal13-2}. However, it is important to keep in mind
that the  limiting magnitude  of the quasar  survey has  been modified
from  $i<$19.1 in  SDSS\,I/II  to $g<$22  in  SDSS\,III. We  therefore
expect a  considerable fraction  of SDSS\,III  WDMS binary  spectra to
generally be  of lower SN  ratio than  those from SDSS\,I/II.   Low SN
ratio spectra  ($\la5$) are  challenging to  classify for  any method,
including our $\chi^2$-SN routine.

\begin{table}
\centering
\caption{\label{t-numbers}  Number  of  WDMS binary  spectra  we  have
  identified or studied (first column),  number of WDMS binary spectra
  added  to  the SDSS\,III  sample  identified  in this  work  (second
  column)  and total  number  of SDSS\,III  WDMS  binary spectra.   In
  brackets  we give  the  number of  WDMS  binary candidate  spectra.}
\setlength{\tabcolsep}{1.8ex}
\begin{tabular}{lccc}
\hline
\hline
 Exercise & N$_\mathrm{sample}$        & N$_\mathrm{added}$  &   N$_\mathrm{SDSS\,III}$ \\ 
\hline 
$\chi^2$-SN                    &   872(47)      &  -          &  872(47) \\ 
WT                             &   1072(30)     &  402(20)    & 1274(67) \\
Completeness                   &   933(26)      &  30(4)      & 1304(71) \\
\citet{lietal14-1}             &   152(4)       &  5(3)       & 1309(74) \\
\citet{kepleretal15-1}         &   158(6)       &  5(4)       & 1314(78) \\
\citet{kepleretal16-1}         &   127(2)       &  5(1)       & 1319(79) \\
\citet{gentilefusilloetal15-1} &   93(0)        &  2(0)       & 1321(79) \\
\hline
WT to SDSS\,I/II               &   1852(13)     & 46(12)      &  -      \\ 
\hline
Total in this work             &                &             & 1367(91) \\
\hline
\end{tabular}
\end{table}

\citet{renetal14-1}  developed an  independent strategy  based on  the
wavelet  transform  \citep[WT,  e.g.][]{chui92-1} that  proved  to  be
efficient   at   identifying   such   low   SN   ratio   WDMS   binary
spectra. \citet{renetal14-1} searched for  WDMS binaries among spectra
obtained  as   part  of   the  LAMOST  (Large   Aperture  Multi-Object
Spectroscopy   Telescope)  surveys   \citep{cuietal12-1,  luoetal12-1,
  liuetal14-1,  yuanetal15-2}.  The  analysis unit  of the  WT is  the
local  flux of  the spectrum,  i.e.  the  selected spectral  features.
Thus, for  a given spectrum,  the WT recognizes the  spectral features
rather than  the global continuum.   The WT decomposes  the considered
features of a given spectrum into approximation signals, also referred
to  as approximation  coefficients.  The  wavelength values  under the
considered spectral regions where the WT is applied are converted into
``data  points''.   The  outcome  of  a  WT  can  be  considered,  for
comparative purposes, as a smooth version  of the spectrum.  The WT is
therefore a suitable method to identify spectral features among low-SN
ratio  spectra, and  we  adopt it  here to  complement  our search  of
SDSS\,III WDMS binaries.

\begin{figure}
\centering
\includegraphics[angle=-90, width=\columnwidth]{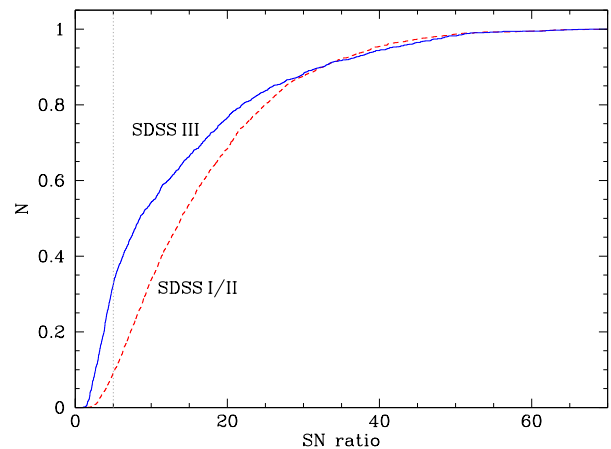}
\caption {Cumulative distributions of SN  ratio of WDMS binary spectra
  identified within  SDSS\,I/II (red dashed line)  and SDSS\,III (blue
  solid line).   The gray  vertical line indicates  SN$=$5.  SDSS\,III
  contains  a  considerably  larger   fraction  of  low  ($\leq$5)  SN
  spectra.}
\label{f-sn}
\end{figure}

We   applied  the   WT  to   the  following   spectral  regions:   the
3910-4422\AA\,  range  thus  covering  the  Balmer  lines  typical  of
hydrogen-rich (DA)  white dwarfs,  and the 6800-8496\AA\,  range which
samples  a large  number  of TiO  and VO  molecular  bands typical  of
low-mass main sequence stars.  This choice implies that our WT routine
is most  efficient at  identifying WDMS  binaries containing  DA white
dwarfs, by  far the most common  among white dwarfs.  Once  we applied
the  WT  to all  SDSS\,III  spectra,  WDMS  binaries could  be  easily
selected  by   applying  a  number   of  cuts  to   the  approximation
coefficients and data points obtained for each spectrum (see Eqs.  (1)
\& (2) of \citealt{renetal14-1}).   Applying the WT and aforementioned
cuts to all SDSS\,III spectra resulted in 36,543 selected spectra.  We
visually inspected the spectra and found  1,072 of these to be genuine
WDMS binaries, and 30 to be WDMS binary candidates.

\section{The SDSS spectroscopic catalogue of WDMS binaries}

Here we quantify the total number  of SDSS\,III WDMS binary spectra we
identified in  the previous  section and determine  how many  of these
objects  are  new additions  to  our  latest  SDSS DR\,8  WDMS  binary
catalogue.  We  also estimate the  completeness of our  SDSS\,III WDMS
binary sample and compare our  catalogue to previously published lists
of SDSS\,III WDMS  binaries.  Finally, we provide the  total number of
systems that form the spectroscopic catalogue of SDSS WDMS binaries.

\subsection{The SDSS\,III WDMS binary sample}

As explained  above, 872  and 47 SDSS\,III  WDMS binary  and candidate
spectra respectively  have been  identified following  the $\chi^2$-SN
ratio method.   Also, 1,072 and  30 WDMS binary and  candidate spectra
were identified following the  WT method (see Table\,\ref{t-numbers}).
Whilst 670  (10 candidates)  are systems  commonly identified  by both
methods, 202 (37 candidates) and 402 (20 candidates) are independently
found by the $\chi^2$-SN ratio  and the WT routines, respectively (see
Figure\,\ref{f-venn},  top panel).   This brings  the total  number of
SDSS\,III  WDMS binary  and candidate  spectra we  have identified  to
1,274 and 67, respectively (Table\,\ref{t-numbers}).

\begin{figure}
\centering
\includegraphics[width=\columnwidth]{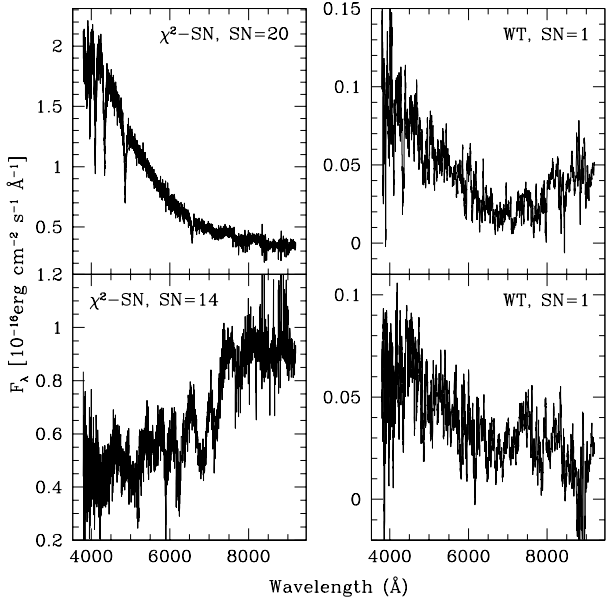}
\caption {Example  WDMS binary spectra  found only by  the $\chi^2$-SN
  method (left panels)  and only by the WT method  (right panels). The
  spectra in the  right panels have been smoothed using  a filter size
  of 15 due to the very low SN ratio. Whilst the $\chi^2$-SN method is
  more efficient  at finding  WDMS binaries  in which  one of  the two
  components dominates the  spectrum, the WT method  is more efficient
  at finding low SN ratio WDMS binary spectra.}
\label{f-spectra}
\end{figure}

Inspection of the numbers outlined  above reveals that the $\chi^2$-SN
ratio method has identified only $\sim$68 per cent of our total number
of  SDSS\,III  WDMS  binary  spectra   and  that  the  WT  method  has
successfully identified  $\sim$84 per cent  of the total  number. That
is, the WDMS  binary samples identified by the $\chi^2$-SN  and the WT
methods are $\sim$68 and $\sim$84 per cent complete, respectively (the
completeness of  the overall sample  is derived in the  next section).
Whilst the  $\sim$84 per cent of  WDMS binary spectra found  by the WT
method is  in agreement  with the  results of  \citet{renetal14-1} ---
they estimate a completeness of $\sim$90 per cent --- the $\sim$68 per
cent of WDMS binary spectra identified by the $\chi^2$-SN ratio method
is considerably  lower than  the $\sim96$ per  cent determined  in our
previous            works            \citep{rebassa-mansergasetal10-1,
  rebassa-mansergasetal12-1,  rebassa-mansergasetal13-2}.  As  we show
in  Figure\,\ref{f-sn},  this seems  to  be  simply a  consequence  of
SDSS\,III WDMS binary spectra being generally of lower SN ratio due to
the limiting magnitude of the BOSS quasar survey \citep{rossetal12-1}.
Indeed, we  investigated the 402  WDMS binary spectra  the $\chi^2$-SN
ratio method failed to identify and found that 81 per cent of them are
of SN$\leq$5.   If we consider  only WDMS  binary spectra of  SN ratio
above 5,  i.e.  a total  of 876 spectra,  then the $\chi^2$-SN  and WT
methods successfully  identify 800 and 677  spectra, respectively, 601
in common (see Figure\,\ref{f-venn}, bottom panel).  That is, $\sim$91
and $\sim$77 per cent of the  total number of WDMS binary spectra with
SN  ratio  $>$5. This  implies  the  WT  method is  considerably  more
efficient at identifying WDMS binary spectra of low SN ratio.  We also
visually  inspected the  list  of  spectra the  WT  methods failed  to
identify and found that it mostly contained WDMS binaries in which the
flux of one of the  two components dominates the spectrum.  Therefore,
we  conclude that  the $\chi^2$-SN  ratio and  WT methods  efficiently
complement each other for  identifying WDMS binaries within SDSS\,III.
In  Figure\,\ref{f-spectra} we  show  a few  examples  of WDMS  binary
spectra identified  only by the  $\chi^2$-SN ratio method and  only by
the WT method.   In the following section we  analyse the completeness
of the entire SDSS\,III WDMS binary sample.

\subsection{Completeness of the SDSS\,III WDMS binary sample}

\begin{table*}
\centering
\caption{\label{t-compl} Combination of WDMS binary stellar parameters
  and  flux  ratios  that  allow  the detection  of  the  two  stellar
  components  in the  SDSS  spectrum.  The  flux  ratios are  directly
  derived from the effective temperatures  of the two components.  The
  effective temperatures of the secondary stars are obtained using the
  empirical   spectral   type-effective   temperature   relation   of
  \citet{camachoetal14-1}.}  \setlength{\tabcolsep}{1ex}
\begin{tabular}{cccccccccc}
\hline
\hline
$\log$\,(g)$_\mathrm{WD}$ & Sp$_\mathrm{sec}$ &  \Teff$_\mathrm{,sec}$ & \Teff$_\mathrm{,WD}$ & Flux Ratio & $\log$\,(g)$_\mathrm{WD}$ & Sp$_\mathrm{sec}$ &  \Teff$_\mathrm{,sec}$ & \Teff$_\mathrm{,WD}$ & Flux Ratio \\
(dex) & & (K) & (K) & & (dex) & & (K) & (K) & \\
\hline
 6.5 &  0 &  4044 &   9000--100000   & 4.08 $\times 10^{-2}$--$2.67 \times 10^{-6}$ &  8.0 &  5 &  3063 &   6000--100000   & 6.79 $\times 10^{-2}$--$8.80 \times 10^{-7}$ \\
 6.5 &  1 &  3738 &   8000--100000   & 4.77 $\times 10^{-2}$--$1.95 \times 10^{-6}$ &  8.0 &  6 &  2947 &   6000--100000   & 5.82 $\times 10^{-2}$--$7.54 \times 10^{-7}$ \\
 6.5 &  2 &  3505 &   7000--100000   & 6.29 $\times 10^{-2}$--$1.51 \times 10^{-6}$ &  8.0 &  7 &  2817 &   6000--70000    & 4.86 $\times 10^{-2}$--$2.62 \times 10^{-6}$ \\
 6.5 &  3 &  3325 &   6000--100000   & 9.43 $\times 10^{-2}$--$1.22 \times 10^{-6}$ &  8.0 &  8 &  2658 &   6000--20000    & 3.85 $\times 10^{-2}$--$3.12 \times 10^{-4}$ \\
 6.5 &  4 &  3184 &   6000--100000   & 7.93 $\times 10^{-2}$--$1.03 \times 10^{-6}$ &  8.0 &  9 &  2453 &   6000--13000    & 2.79 $\times 10^{-2}$--$1.27 \times 10^{-3}$ \\
 6.5 &  5 &  3063 &   6000--75000    & 6.79 $\times 10^{-2}$--$2.78 \times 10^{-6}$ &  8.5 &  0 &  4044 &   20000--100000  & 1.67 $\times 10^{-3}$--$2.67 \times 10^{-6}$ \\
 6.5 &  6 &  2947 &   6000--30000    & 5.82 $\times 10^{-2}$--$9.31 \times 10^{-5}$ &  8.5 &  1 &  3738 &   14000--100000  & 5.08 $\times 10^{-3}$--$1.95 \times 10^{-6}$ \\
 6.5 &  7 &  2903 &   6000--18000    & 5.48 $\times 10^{-2}$--$6.00 \times 10^{-4}$ &  8.5 &  2 &  3505 &   11000--100000  & 1.03 $\times 10^{-2}$--$1.51 \times 10^{-6}$ \\
 6.5 &  8 &  2658 &   6000--9000     & 3.85 $\times 10^{-2}$--$7.61 \times 10^{-3}$ &  8.5 &  3 &  3325 &   9000--100000   & 1.86 $\times 10^{-2}$--$1.22 \times 10^{-6}$ \\
 6.5 &  9 &  2453 &   6000--7000     & 2.79 $\times 10^{-2}$--$1.51 \times 10^{-2}$ &  8.5 &  4 &  3184 &   8000--100000   & 2.51 $\times 10^{-2}$--$1.03 \times 10^{-6}$ \\
 7.0 &  0 &  4044 &   9000--100000   & 4.08 $\times 10^{-2}$--$2.67 \times 10^{-6}$ &  8.5 &  5 &  3063 &   7000--100000   & 3.67 $\times 10^{-2}$--$8.80 \times 10^{-7}$ \\
 7.0 &  1 &  3738 &   8000--100000   & 4.77 $\times 10^{-2}$--$1.95 \times 10^{-6}$ &  8.5 &  6 &  2947 &   6000--100000   & 5.82 $\times 10^{-2}$--$7.54 \times 10^{-7}$ \\
 7.0 &  2 &  3505 &   7000--100000   & 6.29 $\times 10^{-2}$--$1.51 \times 10^{-6}$ &  8.5 &  7 &  2817 &   6000--100000   & 4.86 $\times 10^{-2}$--$6.30 \times 10^{-7}$ \\
 7.0 &  3 &  3325 &   6000--100000   & 9.43 $\times 10^{-2}$--$1.22 \times 10^{-6}$ &  8.5 &  8 &  2658 &   6000--36000    & 3.85 $\times 10^{-2}$--$2.97 \times 10^{-5}$ \\
 7.0 &  4 &  3184 &   6000--100000   & 7.93 $\times 10^{-2}$--$1.03 \times 10^{-6}$ &  8.5 &  9 &  2453 &   6000--20000    & 2.79 $\times 10^{-2}$--$2.26 \times 10^{-4}$ \\
 7.0 &  5 &  3063 &   6000--100000   & 6.79 $\times 10^{-2}$--$8.80 \times 10^{-7}$ &  9.0 &  0 &  4044 &   30000--100000  & 3.30 $\times 10^{-4}$--$2.67 \times 10^{-6}$ \\
 7.0 &  6 &  2947 &   6000--55000    & 5.82 $\times 10^{-2}$--$8.24 \times 10^{-6}$ &  9.0 &  1 &  3738 &   24000--100000  & 5.88 $\times 10^{-4}$--$1.95 \times 10^{-6}$ \\
 7.0 &  7 &  2817 &   6000--24000    & 4.86 $\times 10^{-2}$--$1.90 \times 10^{-4}$ &  9.0 &  2 &  3505 &   18000--100000  & 1.44 $\times 10^{-3}$--$1.51 \times 10^{-6}$ \\
 7.0 &  8 &  2658 &   6000--10000    & 3.85 $\times 10^{-2}$--$4.99 \times 10^{-3}$ &  9.0 &  3 &  3325 &   12000--100000  & 5.89 $\times 10^{-3}$--$1.22 \times 10^{-6}$ \\
 7.0 &  9 &  2453 &   6000--8000     & 2.79 $\times 10^{-2}$--$8.84 \times 10^{-3}$ &  9.0 &  4 &  3184 &   10000--100000  & 1.03 $\times 10^{-2}$--$1.03 \times 10^{-6}$ \\
 7.5 &  0 &  4044 &   11000--100000  & 1.83 $\times 10^{-2}$--$2.67 \times 10^{-6}$ &  9.0 &  5 &  3063 &   9000--100000   & 1.34 $\times 10^{-2}$--$8.80 \times 10^{-7}$ \\
 7.5 &  1 &  3738 &   9000--100000   & 2.98 $\times 10^{-2}$--$1.95 \times 10^{-6}$ &  9.0 &  6 &  2947 &   7000--100000   & 3.14 $\times 10^{-2}$--$7.54 \times 10^{-7}$ \\
 7.5 &  2 &  3505 &   8000--100000   & 3.68 $\times 10^{-2}$--$1.51 \times 10^{-6}$ &  9.0 &  7 &  2817 &   6000--100000   & 4.86 $\times 10^{-2}$--$6.30 \times 10^{-7}$ \\
 7.5 &  3 &  3325 &   7000--100000   & 5.09 $\times 10^{-2}$--$1.22 \times 10^{-6}$ &  9.0 &  8 &  2658 &   6000--95000    & 3.85 $\times 10^{-2}$--$6.13 \times 10^{-7}$ \\
 7.5 &  4 &  3184 &   6000--100000   & 7.93 $\times 10^{-2}$--$1.03 \times 10^{-6}$ &  9.0 &  9 &  2453 &   6000--30000    & 2.79 $\times 10^{-2}$--$4.47 \times 10^{-5}$ \\
 7.5 &  5 &  3063 &   6000--100000   & 6.79 $\times 10^{-2}$--$8.80 \times 10^{-7}$ &  9.5 &  1 &  3738 &   60000--100000  & 1.51 $\times 10^{-5}$--$1.95 \times 10^{-6}$ \\
 7.5 &  6 &  2947 &   6000--95000    & 5.82 $\times 10^{-2}$--$9.26 \times 10^{-7}$ &  9.5 &  2 &  3505 &   30000--100000  & 1.86 $\times 10^{-4}$--$1.51 \times 10^{-6}$ \\
 7.5 &  7 &  2817 &   6000--40000    & 4.86 $\times 10^{-2}$--$2.46 \times 10^{-5}$ &  9.5 &  3 &  3325 &   22000--100000  & 5.22 $\times 10^{-4}$--$1.22 \times 10^{-6}$ \\
 7.5 &  8 &  2658 &   6000--16000    & 3.85 $\times 10^{-2}$--$7.62 \times 10^{-4}$ &  9.5 &  4 &  3184 &   18000--100000  & 9.79 $\times 10^{-4}$--$1.03 \times 10^{-6}$ \\
 7.5 &  9 &  2453 &   6000--10000    & 2.79 $\times 10^{-2}$--$3.62 \times 10^{-3}$ &  9.5 &  5 &  3063 &   14000--100000  & 2.29 $\times 10^{-3}$--$8.80 \times 10^{-7}$ \\
 8.0 &  0 &  4044 &   15000--100000  & 5.28 $\times 10^{-3}$--$2.67 \times 10^{-6}$ &  9.5 &  6 &  2947 &   10000--100000  & 7.54 $\times 10^{-3}$--$7.54 \times 10^{-7}$ \\
 8.0 &  1 &  3738 &   11000--100000  & 1.33 $\times 10^{-2}$--$1.95 \times 10^{-6}$ &  9.5 &  7 &  2817 &   8000--100000   & 1.54 $\times 10^{-2}$--$6.30 \times 10^{-7}$ \\
 8.0 &  2 &  3505 &   9000--100000   & 2.30 $\times 10^{-2}$--$1.51 \times 10^{-6}$ &  9.5 &  8 &  2658 &   7000--100000   & 2.08 $\times 10^{-2}$--$1.51 \times 10^{-2}$ \\
 8.0 &  3 &  3325 &   8000--100000   & 2.98 $\times 10^{-2}$--$1.22 \times 10^{-6}$ &  9.5 &  9 &  2453 &   10000--100000  & 3.62 $\times 10^{-3}$--$3.62 \times 10^{-7}$ \\
 8.0 &  4 &  3184 &   7000--100000   & 4.28 $\times 10^{-2}$--$1.03 \times 10^{-6}$ &      &    &       &                  &                                            \\
\hline
\end{tabular}
\end{table*}

Using the $\chi^{2}$-SN  and WT methods we have identified  a total of
1,274   and   67  SDSS\,III   WDMS   binary   and  candidate   spectra
(Table\,\ref{t-numbers}).  Here,  we analyse how complete  this sample
is.

\begin{figure}
\centering
\includegraphics[angle=-90, width=\columnwidth]{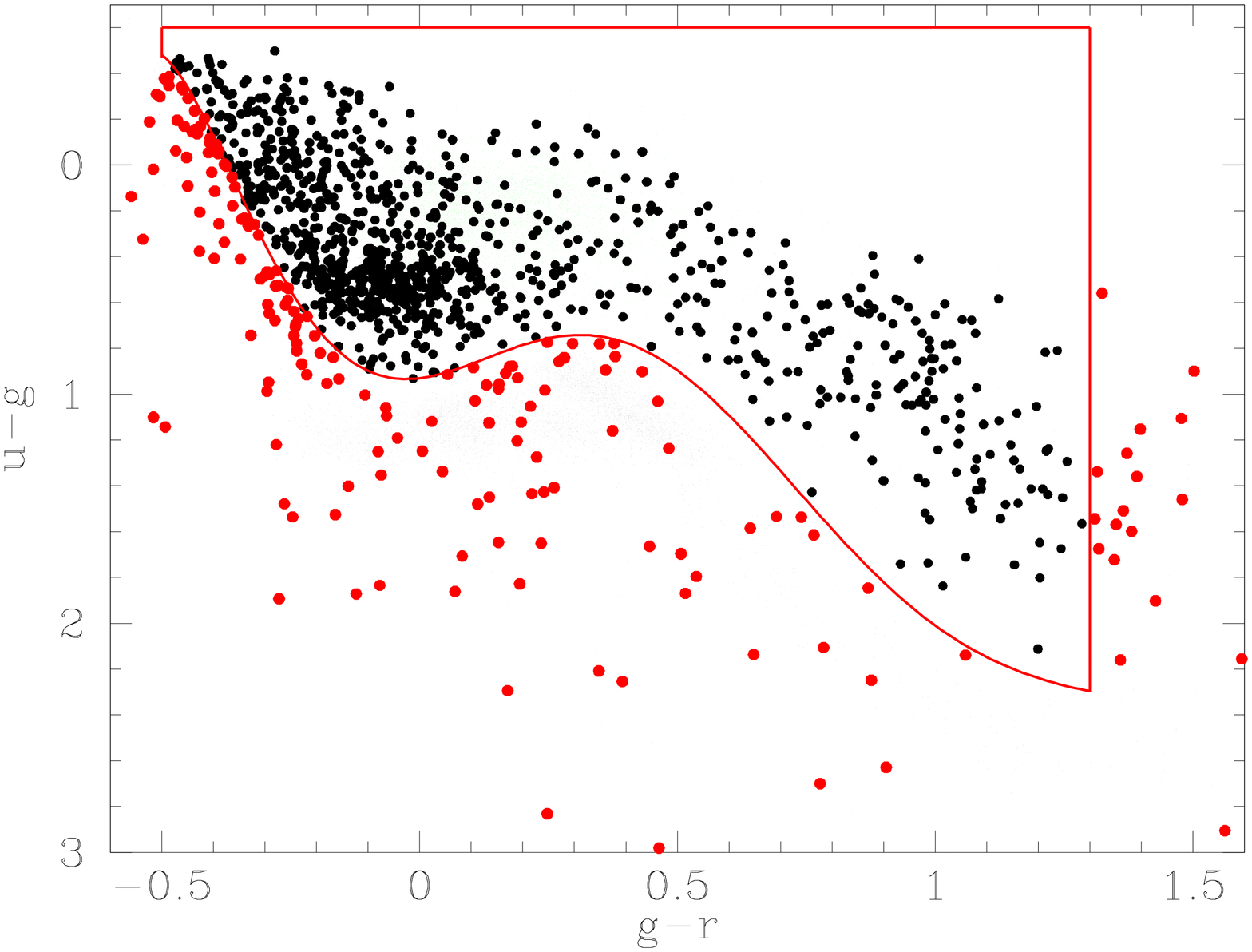}
\caption {SDSS\,III WDMS binaries  identified by the $\chi^{2}$-SN and
  WT methods (black  solid dots), main sequence stars  (gray dots) and
  quasars  (green solid  dots) in  the $u-g$  vs. $g-r$  colour plane.
  Colour  cuts  for  WDMS  binaries  are given  as  solid  red  lines.
  $\sim$15  per cent  of all  WDMS  binaries fall  outside the  colour
  selection mainly due to the large photometric uncertainties.}
\label{f-cuts}
\end{figure}

WDMS  binaries   define  a  clear   region  in  the  $u-g$   vs  $g-r$
colour-colour diagram \citep{smolcicetal04-1} and  can be selected via
the     colour     cuts      provided     by     \citet[][see     also
  Figure\,\ref{f-cuts}]{rebassa-mansergasetal13-2}:

\begin{eqnarray}
(u - g)< 0.93-0.27\times(g - r)-4.7\times(g - r)^{2} \nonumber\\ 
+ 12.38\times(g - r)^{3}+3.08\times(g - r)^{4}-22.19\times(g - r)^{5}\nonumber \\
 +16.67\times(g - r)^{6} - 3.89\times(g - r)^{7}, \\
(u - g)>-0.6, \\
-0.5<(g - r)<1.3
\end{eqnarray}

Obviously, the efficiency  of the above colour cuts  in selecting WDMS
binaries  relies on  the accuracy  of  the SDSS  photometry.  This  is
difficult to achieve  in cases where the two  components are partially
resolved (blended) in the SDSS images,  when the systems are too faint
so that their magnitudes are associated to large uncertainties, and/or
when one of  the the two stars outshines its  companion.  This results
in $\sim$15 per cent of the  SDSS\,III WDMS binaries identified by the
$\chi^{2}$-SN  and WT  methods to  fall outside  the colour  cuts (see
Figure\,\ref{f-cuts}).  However, it has to be stressed that the colour
selection can  be used to  test the completeness of  our spectroscopic
sample,  as  visual inspection  of  all  sources with  available  SDSS
spectra satisfying the cuts would result  in a WDMS binary sample that
is 100 per cent complete (within the colour selection).  We hence used
the $casjobs$ interface to select all spectroscopic point sources with
clean  photometry satisfying  the  cuts.  This  resulted in  1,264,475
objects among  7,121,388 photometric  sources.  Excluding  all targets
with  spectroscopic  redshifts  below   0.05  (to  thus  avoid  quasar
contamination) we ended up with  50,691 SDSS\,III selected objects (in
this exercise  we did not take  into account all spectra  contained in
the  DR\,8  or earlier  SDSS  releases).   We visually  inspected  the
spectra  and  identified  933  (26) WDMS  binary  (candidate)  spectra
(Table\,\ref{t-numbers}).   Of these,  only  30 (4)  spectra were  not
included in our list of systems identified by the $\chi^{2}$-SN and WT
methods.   Visual inspection  of  the spectra  we  failed to  identify
revealed that, in most of the  cases, one of the two binary components
dominates the  spectral energy  distribution.  This  exercise strongly
suggests  that our  SDSS\,III WDMS  binary sample  is highly  complete
($\simeq$96 per  cent) and that  the WDMS  binary spectra we  miss are
likely dominated  by the  flux of  one of  the binary  components.  We
include the  34 missed  spectra in  our sample  and thus  increase the
number of SDSS\,III WDMS binary and candidate spectra to 1,304 and 71,
respectively (Table\,\ref{t-numbers}).

\begin{figure}
\centering
\includegraphics[angle=-90, width=\columnwidth]{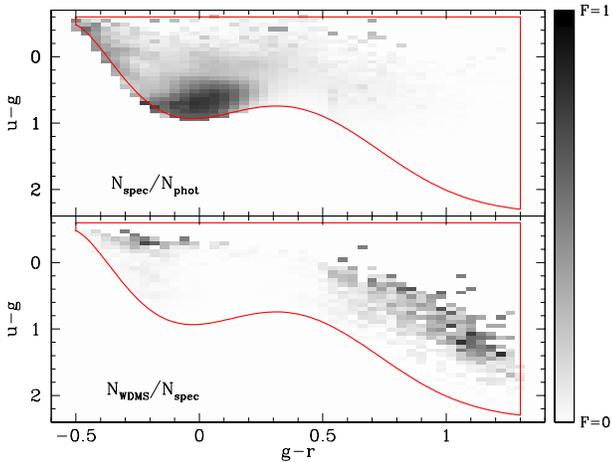}
\caption {Top  panel: density  map illustrating  the fraction  of SDSS
  objects            with            available            spectroscopy
  (F=N$_\mathrm{spec}$/N$_\mathrm{phot}$, ranging  from 0 to  1, where
  N$_\mathrm{spec}$  and  N$_\mathrm{phot}$  are the  number  of  SDSS
  objects with  available spectra and  the number of  SDSS photometric
  point  sources, respectively).   As  expected, the  fraction is  the
  highest  in  quasar  dominated  areas.  Bottom  panel:  density  map
  illustrating  the  fraction  of  WDMS binaries  among  objects  with
  available    spectroscopy    (F=N$_\mathrm{WDMS}$/N$_\mathrm{spec}$,
  ranging from 0  to 1, where N$_\mathrm{WDMS}$ is the  number of SDSS
  WDMS  binaries).  The  fraction is  the lowest  in quasar  dominated
  areas due to the large number of observed quasars.}
\label{f-density}
\end{figure}

It is important  to emphasise that, although  our spectroscopic sample
is  highly complete,  the population  of SDSS  WDMS binaries  does not
represent the  real and unbiased  population of WDMS  binaries. First,
there exists  an intrinsic binary bias  that allows us to  detect only
those  objects displaying  both components  in the  SDSS spectra.   We
studied this  issue in  detail in  \citet{camachoetal14-1} so  we here
provide in Table\,\ref{t-compl} the  combination of stellar parameters
and flux ratios that allow the  detection of the two components in the
SDSS spectra.   Second, the majority  of SDSS WDMS binaries  have been
observed simply because they have similar colours to those of quasars.
This effect is clearly visible in Figure\,\ref{f-cuts}, where there is
a clear scarcity  of WDMS systems with $g-r>0.5$ colours,  and also in
Figure\,\ref{f-density}  (top  panel), where  we  show  a density  map
illustrating  the fraction  of  SDSS objects  with available  spectra.
This implies the majority of spectroscopic SDSS WDMS binaries generally
contain    hot    white    dwarfs     (see    further    details    in
Section\,\ref{s-comparison}).

\subsection{Comparison with \citet{lietal14-1}}

\citet{lietal14-1}  provided a  list of  227 DR\,9  WDMS binaries,  of
which 148  (1) are included in  our WDMS binary (candidate)  list.  Of
the remaining 78, 70 spectra are not considered as WDMS binaries by us
(the majority are main sequence  plus main sequence superpositions and
also some single white dwarfs), and  5 (3) are WDMS binary (candidate)
spectra that we missed.  This  implies we have successfully identified
$\sim$95   per   cent   of   the   total   WDMS   binary   sample   of
\citet{lietal14-1}.   The missed  spectra  are dominated  by the  flux
emission  of one  of the  two components  and we  include them  in our
lists, thus raising the number  of SDSS\,III WDMS binary and candidate
spectra to  1,309 and  74, respectively  (Table\,\ref{t-numbers}).  We
also note that our list contains 447 WDMS binary spectra that are part
of  SDSS   DR\,9  which   are  not  included   in  the   catalogue  of
\citet{lietal14-1}.  This  indicates that  our routines are  much more
efficient at identifying SDSS WDMS binaries, and that the catalogue of
SDSS WDMS binaries by \citet{lietal14-1} is highly incomplete.

\citet{lietal14-1} provide  effective temperatures,  surface gravities
and masses  for the white dwarfs  in their WDMS binary  sample.  As we
also do, they derive these  parameters fitting the white dwarf spectra
with the 1D  white dwarf model grid  of \citet{koester10-1}.  However,
they  do not  apply 3D  corrections to  avoid the  so-called $\log  g$
problem  at  low  effective   temperatures  (see  further  details  in
Section\,\ref{s-param}).

\begin{figure*}
\centering
\includegraphics[angle=-90, width=\textwidth]{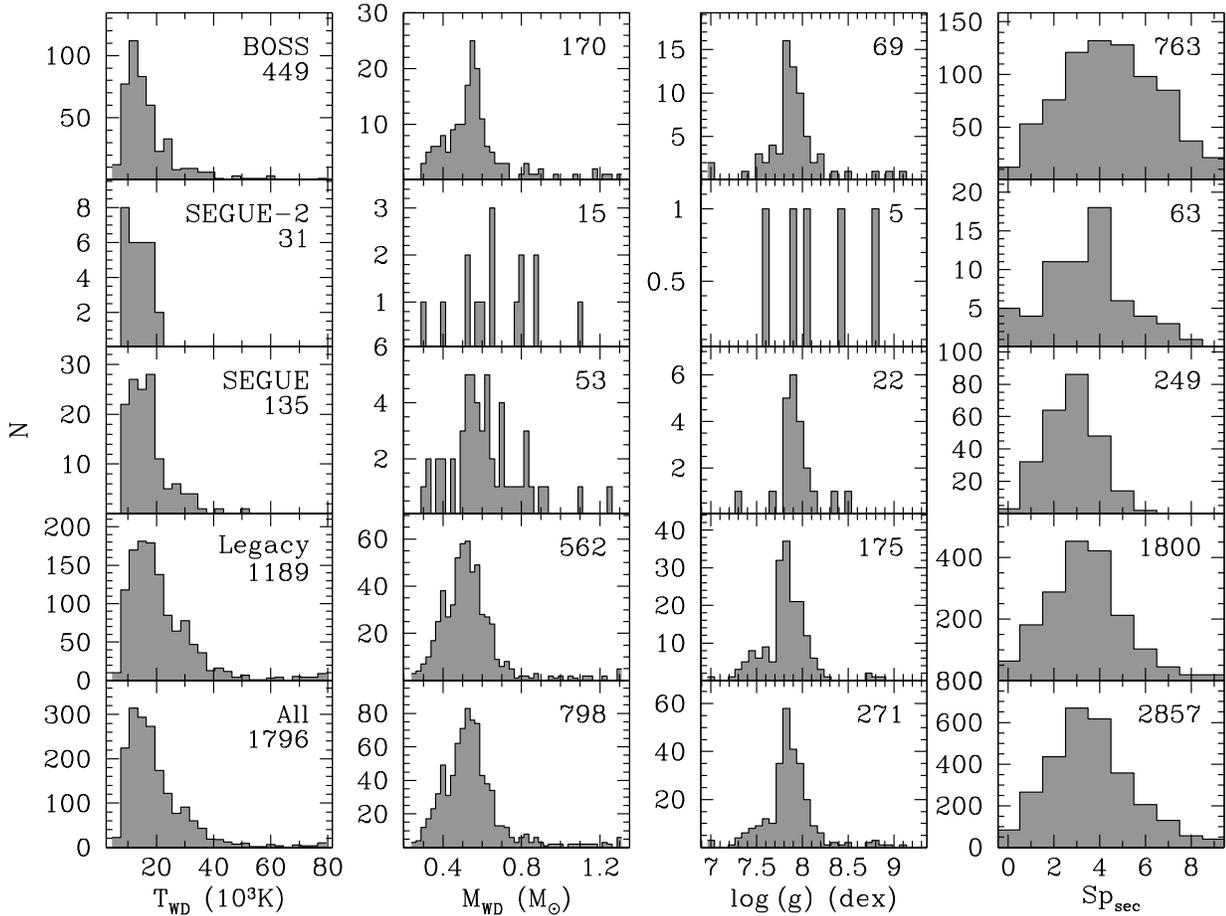}
\caption {From left to right the distribution of white dwarf effective
  temperatures,  masses,  surface  gravities and  secondary  star  (M)
  spectral types for the WDMS  binaries observed by the BOSS, SEGUE-2,
  SEGUE and Legacy  surveys of SDSS.  Given that WDMS  binaries can be
  observed by more than one survey, we  consider them to be part of as
  many samples as different surveys  have observed them.  In the bottom
  panels   we  provide   the  distributions   for  the   entire  (All)
  spectroscopic catalogue of SDSS DR\,12  WDMS binaries.  In this case
  WDMS binaries  that are observed  by more  than one survey  are only
  considered  once.   We  also  provide  the  number  of  systems  per
  distribution in each panel.}
\label{f-histo}
\end{figure*}

\subsection{Comparison with \citet{kepleretal15-1} and \citet{kepleretal16-1}}

An additional list  of 177 DR\,9-10 SDSS WDMS binary  spectra has been
provided by \citet{kepleretal15-1}.  153 (2)  of these are included in
our WDMS  binary (candidate) list,  we only  missed 5 (4)  WDMS binary
(candidate) spectra  (Table\,\ref{t-numbers}), the  majority dominated
by  the  flux  contribution  of  the white  dwarf.   We  classify  the
remaining 13 spectra as single white dwarfs or main sequence plus main
sequence  superpositions.   Hence,  we  have  successfully  identified
$\sim$94    per   cent    of    the   total    sample   provided    by
\citet{kepleretal15-1}.  We include the 9  missed objects in our lists
and increase to  1,314 and 78 the number of  WDMS binary and candidate
spectra,  respectively  (Table\,\ref{t-numbers}).  Moreover,  we  note
that we have identified 759 DR\,9-10  WDMS binary spectra that are not
included  in  the  list  of  \citet{kepleretal15-1}.   The  fact  that
\citet{kepleretal15-1} missed this large number of WDMS binary spectra
is not  too surprising,  as their  main focus  was to  identify single
white dwarfs.

\citet{kepleretal16-1} presented an updated SDSS white dwarf catalogue
from DR\,12, which includes 160 new WDMS binary spectra.  123 of these
are included  in our lists,  122 as genuine WDMS  binaries and 1  as a
candidate (Table\,\ref{t-numbers}).  The  remaining 37 spectra include
5 (1) WDMS binaries (candidate) that  we missed and 31 objects that we
classify  as  either single  white  dwarfs,  cataclysmic variables  or
unclassified.   Thus, we  have identified  $\sim$95 per  cent of  WDMS
binary sample  provided by  \citet{kepleretal16-1}.  We include  the 6
missed objects in our lists thus increasing to 1,319 and 79 the number
of    WDMS     binary    and    candidate     spectra,    respectively
(Table\,\ref{t-numbers}).  We  also note  that our  catalogue contains
337  DR\,12 WDMS  binary spectra  that \citet{kepleretal16-1}  did not
identify.

\subsection{Comparison with \citet{gentilefusilloetal15-1}}

\citet{gentilefusilloetal15-1} developed  a method which uses  cuts in
colour-colour  and  reduced  proper   motion-colour  space  to  select
$\sim$23\,000 high-fidelity white dwarf candidates. Among the 8,701 of
them with available  spectra, they classified 98 as  WDMS binaries, of
which 91 are  included in our WDMS binary catalogue.   The remaining 7
include 2 WDMS  binaries that we missed clearly dominated  by the flux
contribution of  the white dwarfs  and 5  objects that we  classify as
single white dwarfs or cataclysmic variables.  We thus have identified
99     per     cent    of     the     WDMS     binary    sample     of
\citet{gentilefusilloetal15-1}.  We  include the  2 missed  objects in
our list and thus raise to 1,321  and 79 the number of WDMS binary and
candidate spectra, respectively (Table\,\ref{t-numbers}).

\begin{table}
\centering
\caption{\label{t-numbers2}  Number   of  SDSS  WDMS   binary  spectra
  identified  in this  work (N$_\mathrm{spectra}$),  number of  unique
  sources  after excluding  duplicated spectra  (N$_\mathrm{unique}$),
  number  of additions  to  our latest  DR\,8 spectroscopic  catalogue
  after    excluding    systems    that    were    already    observed
  (N$_\mathrm{additions}$), total number  of spectroscopic SDSS DR\,12
  WDMS binaries  (N$_\mathrm{total}$) and  number of new  systems that
  have not  been published before (N$_\mathrm{new}$).   In brackets we
  give     the     number      of     WDMS     binary     candidates.}
\setlength{\tabcolsep}{2.8ex}
\begin{tabular}{cc}
\hline
\hline
N$_\mathrm{spectra}$         &  1367\,(91)   \\
N$_\mathrm{unique}$          &  1177\,(91)   \\
N$_\mathrm{additions}$       &  892\,(87)    \\
N$_\mathrm{total}$           &  3115\,(180)   \\
N$_\mathrm{new}$             &  572\,(75)    \\
\hline
\end{tabular}
\end{table}

\subsection{Applying the WT method to DR\,8 spectra}

We  have shown  that the  WT is  more efficient  than the  $\chi^2$-SN
method  at identifying  SDSS WDMS  binaries  of low  SN ratio  spectra
(Section\,\ref{s-ident}).     It   is    therefore   plausible    that
\citet{rebassa-mansergasetal10-1,           rebassa-mansergasetal12-1,
  rebassa-mansergasetal13-2}  missed  a  fraction   of  low  SN  ratio
SDSS\,I/II  WDMS binary  spectra.   In order  to  investigate this  we
applied  the WT  as specified  in Section\,\ref{s-ident}  to all  SDSS
DR\,8 spectra, which resulted in 20,060 selected candidates.  We found
1,852 to be genuine WDMS binaries and 13 are considered as WDMS binary
candidates.  Only the 12 candidate  spectra and 46 WDMS binary spectra
are    not   included    in   our    SDSS   WDMS    binary   catalogue
(Table\,\ref{t-numbers}) --- $\sim$3 per  cent of the total identified
sample \citep{rebassa-mansergasetal13-2}.

\subsection{The final number of spectroscopic SDSS WDMS binaries}

In the previous sections we have identified a total of 1,367 (91) WDMS
binary     (candidate)    spectra     (Table\,\ref{t-numbers}).     We
cross-correlate this  list with  our latest  catalogue of  2,316 DR\,8
WDMS  binaries  \citep{rebassa-mansergasetal13-2}  and  find  979  new
systems,   647    of   which   have   not    been   published   before
(Table\,\ref{t-numbers2}).   Thus, the  total number  of spectroscopic
SDSS WDMS binaries raises to 3,295,  i.e.  an increase of $\sim$40 per
cent.  We  have updated  our web  site \emph{http://www.sdss-wdms.org}
including the  new systems and  spectra identified in this  work.  The
object  names, coordinates  and $ugriz$  magnitudes of  the SDSS  WDMS
binaries  found in  this work  are  also available  in the  electronic
edition of the paper.

\section{Stellar parameters}
\label{s-param}

We here  derive white dwarf effective  temperatures, surface gravities
and masses,  and secondary star  spectral types for all  WDMS binaries
identified in this work. To  that end we use the decomposition/fitting
routine outlined by \citet{rebassa-mansergasetal07-1}.  We first use a
combined set  of observed M dwarf  and white dwarf templates  to fit a
given SDSS  WDMS binary spectrum and  record the spectral type  of the
secondary star.   We then subtract  the best-fit M dwarf  template and
fit the normalized  Balmer lines of the residual  white dwarf spectrum
with a model grid of DA white dwarfs \citep{koester10-1} to derive the
effective temperature and surface gravity.  The Balmer line fitting is
subject to a ``cold/hot'' solution degeneracy, which is broken fitting
the entire spectrum (continuum plus lines) with the same grid of model
spectra (note  that the continuum  is most sensitive to  the effective
temperature).  Given that  1D white dwarf model spectra  such as those
used in this work yield overestimated surface gravity values for white
dwarfs    of     effective    temperatures     below    $\sim$12000\,K
\citep[e.g.][]{koesteretal09-1,  tremblayetal11-1},  we apply  the  3D
corrections of  \citet{tremblayetal13-1} to our  effective temperature
and surface gravity determinations (we apply the 3D corrections to all
SDSS  WDMS binaries  we identified  in  previous works  too, as  these
corrections were  not available  at that  time).  We  then interpolate
these    values    in    the     updated    cooling    sequences    of
\citet{bergeronetal95-2} and derive white  dwarf masses.  For the sake
of comparison, we also derive the  white dwarf masses using the models
of \citet{renedoetal10-1}  and find no substantial  difference between
the values obtained from the  two cooling sequences.  The SDSS spectra
of the WDMS binaries for which we cannot obtain stellar parameters are
either too noisy and/or are dominated by  the flux from one of the two
stellar components.  In cases where  more than one spectrum per target
is available we average the  corresponding parameter values. The white
dwarf  effective  temperatures,  surface  gravities  and  masses,  and
secondary star spectral types that  result from fitting spectra of the
same targets are found to agree at the 1.5$\sigma$ level in 53, 64, 67
and 89  per cent of  the cases, respectively.  The  stellar parameters
are accessible via our web site \emph{http://www.sdss-wdms.org}.

The parameter  distributions are  shown Fig.\,\ref{f-histo},  where we
divide  the sample  into BOSS  (top  panels) and  SEGUE-2 (second  top
panels) WDMS  binaries.  In the  distributions we only  consider white
dwarf effective temperatures with a  relative error under 10 per cent,
and white dwarf surface gravities  and masses of absolute errors under
0.075\,\Msun\, and 0.075 dex,  respectively.  For completeness, in the
mid-bottom panels  of the  same Figure we  also provide  the parameter
distributions  of SDSS\,I/II  WDMS binaries,  where these  are divided
into     objects    that     were    observed     by    the     Legacy
\citep{adelman-mccarthyetal08-1, abazajianetal09-1}  survey of quasars
and galaxies  (hereafter Legacy WDMS  binaries) and objects  that were
observed as part of our  dedicated SEGUE \citep[the SDSS Extension for
  Galactic Understanding and Exploration,][]{yannyetal09-1} survey for
targeting  objects  containing  cool   white  dwarfs  and  early  type
companions \citep{rebassa-mansergasetal12-1}.  We  restrict the Legacy
and SEGUE samples by the same  error parameter cuts as above outlined.
It is  also important  to mention  that we  do not  exclude duplicated
targets  that have  been  observed by  different  surveys.  Thus,  for
example,  a given  WDMS binary  is  part of  both the  SEGUE and  BOSS
populations if  it has  been independently  observed by  both surveys.
Finally, in the  bottom panel of Figure\,\ref{f-histo}  we provide the
parameter distributions of the  entire spectroscopic catalogue of SDSS
DR\,12 WDMS  binaries, i.e.  the  sum of all WDMS  binary sub-samples.
In this case we do exclude  duplicated targets that have been observed
by  different  surveys.   In  the following  section  we  compare  the
different SDSS WDMS binary populations.

\begin{figure}
\centering
\includegraphics[angle=-90, width=\columnwidth]{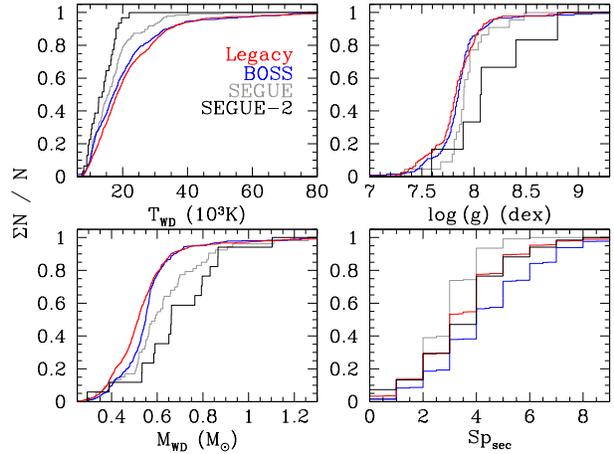}
\caption   {Cumulative   distributions   of  white   dwarf   effective
  temperature  (top left),  mass (bottom  left), surface  gravity (top
  right) and secondary star spectral type (bottom right) for the BOSS,
  SEGUE-2, Legacy and SEGUE WDMS binary populations.}
\label{f-ks}
\end{figure}

\section{A comparison of the different SDSS WDMS binary populations}
\label{s-comparison}

SDSS WDMS  binaries come in different  flavours. On the one  hand, the
majority of  WDMS binaries  are observed by  SDSS simply  because they
have similar colours  as quasars. These WDMS  binaries include objects
observed by  the Legacy survey  of SDSS\,I/II and objects  observed by
the BOSS survey  of SDSS\,III.  On the  other hand, as can  be seen in
Figure\,\ref{f-histo}, a small number of  WDMS binaries have been also
observed by the SEGUE-2 survey of SDSS\,III.  Finally, SDSS\,I/II WDMS
binaries were  additionally observed  as part  of our  dedicated SEGUE
survey  \citep{rebassa-mansergasetal12-1}.   Here   we  compare  these
different  SDSS  WDMS  binary   populations.   The  stellar  parameter
cumulative  distributions  of   white  dwarf  effective  temperatures,
surface gravities  and masses, and  secondary star spectral  types for
the different WDMS binary populations are shown in Figure\,\ref{f-ks}.

\begin{table}
\centering
\caption{\label{t-probs} Kolmogorov-Smirnov and $\chi^2$ probabilities
  that   result   from    comparing   the   cumulative   distributions
  (Figure\,\ref{f-ks}) of the different SDSS WDMS binary populations.}
\setlength{\tabcolsep}{1.8ex}
\begin{tabular}{lccc}
\hline
\hline
 \Twd   & SEGUE-2              & SEGUE              & Legacy              \\ 
BOSS    &  $4 \times 10^{-4}$   & $3 \times 10^{-3}$  & $3 \times 10^{-4}$   \\ 
SEGUE-2 &                      & 0.06               & $2 \times 10^{-5}$   \\ 
SEGUE   &                      &                    & $2 \times 10^{-5}$   \\ 
\hline 
\Mwd    & SEGUE-2              & SEGUE              & Legacy              \\
BOSS    & $2 \times 10^{-4}$    & $3 \times 10^{-3}$  & $9 \times 10^{-7}$   \\
SEGUE-2 &                      & 0.19               & $1 \times 10^{-4}$   \\
SEGUE   &                      &                    & $2 \times 10^{-4}$   \\
\hline
$\log$ g& SEGUE-2              & SEGUE              & Legacy              \\
BOSS    &  0.03                & 0.05               & 0.11                \\
SEGUE-2 &                      & 0.15               & 0.03                \\
SEGUE   &                      &                    & 0.02                \\
\hline
 Sp     & SEGUE-2              & SEGUE              & Legacy              \\
BOSS    & 0.01                 &  0                 & 0                   \\
SEGUE-2 &                      &  0                 & 0.55                \\
SEGUE   &                      &                    & 0                   \\
\hline
\end{tabular}
\end{table}

\begin{table*}
\centering
\caption{\label{t-rvs} 62 new SDSS PCEBs identified in this work.  The
  \Ion{Na}{I} and  H$\alpha$ columns indicate  whether (1) or  not (0)
  the system  displays more  than 3$\sigma$ \Ion{Na}{I}  and H$\alpha$
  emission RV  variation, respectively. WDMS binaries  displaying only
  H$\alpha$  RV variation  should  be considered  as PCEB  candidates,
  except the eclipser SDSSJ093947.95+325807.3 (Section\,\ref{s-eclip})
  and    SDSSJ125645.47+252241.6    (see    Figure\,\ref{f-periods}).}
\setlength{\tabcolsep}{0.5ex}
\begin{small}
\begin{tabular}{cccccccccccc}
\hline
\hline
object & \Ion{Na}{I} & H$\alpha$ & object & \Ion{Na}{I} & H$\alpha$ & object & \Ion{Na}{I} & H$\alpha$& object & \Ion{Na}{I} & H$\alpha$ \\
\hline
  SDSSJ002547.09+134129.8 & 0 & 1 &  SDSSJ075356.37+233118.9 & 1 & 0 &  SDSSJ101642.94+044317.7 & 0 & 1 &  SDSSJ132040.27+661214.8 & 1 & 0 \\
  SDSSJ002926.82+252553.9 & 0 & 1 &  SDSSJ075335.00+510605.2 & 1 & 1 &  SDSSJ101739.89+594829.6 & 0 & 1 &  SDSSJ132830.92+125941.4 & 1 & 0 \\
  SDSSJ003239.97-002611.1 & 1 & 0 &  SDSSJ075835.11+482523.6 & 1 & 1 &  SDSSJ101954.64+531736.9 & 1 & 0 &  SDSSJ134100.03+602610.4 & 1 & 1 \\
  SDSSJ003336.49+004151.3 & 1 & 1 &  SDSSJ080003.86+503545.2 & 0 & 1 &  SDSSJ102750.05+271244.1 & 1 & 1 &  SDSSJ135039.54+261511.7 & 0 & 1 \\
  SDSSJ003402.18+325506.9 & 1 & 0 &  SDSSJ080235.69+525736.3 & 1 & 0 &  SDSSJ103008.34+393715.4 & 0 & 1 &  SDSSJ140138.47+121736.0 & 1 & 1 \\
  SDSSJ010428.86-000907.3 & 1 & 1 &  SDSSJ081409.82+531921.3 & 1 & 1 &  SDSSJ104606.35+583907.8 & 1 & 1 &  SDSSJ142522.83+210940.6 & 1 & 0 \\
  SDSSJ011845.09+303345.2 & 1 & 0 &  SDSSJ081942.67+542608.1 & 0 & 1 &  SDSSJ113829.09+202040.2 & 1 & 0 &  SDSSJ145443.65+570139.6 & 0 & 1 \\
  SDSSJ014745.02-004911.1 & 1 & 1 &  SDSSJ084028.85+501238.2 & 1 & 1 &  SDSSJ114349.96+501020.2 & 1 & 1 &  SDSSJ150605.75+265457.4 & 1 & 1 \\
  SDSSJ020756.15+214027.4 & 1 & 1 &  SDSSJ085515.48+280211.8 & 0 & 1 &  SDSSJ115411.77+543251.4 & 0 & 1 &  SDSSJ154145.39+412230.4 & 1 & 0 \\
  SDSSJ022405.97+002100.6 & 0 & 1 &  SDSSJ090451.86+453105.0 & 0 & 1 &  SDSSJ123234.07+601629.9 & 1 & 1 &  SDSSJ160553.66+085954.4 & 1 & 1 \\
  SDSSJ022436.80+005814.3 & 1 & 0 &  SDSSJ093155.66+394607.7 & 1 & 1 &  SDSSJ123406.86+443115.2 & 0 & 1 &  SDSSJ163831.67+292236.8 & 1 & 0 \\
  SDSSJ024310.60+004044.4 & 1 & 0 &  SDSSJ093947.95+325807.3 & 0 & 1 &  SDSSJ123642.17+005448.3 & 0 & 1 &  SDSSJ212320.23+054253.6 & 1 & 1 \\
  SDSSJ025301.60-013006.9 & 0 & 1 &  SDSSJ094853.94+573957.7 & 0 & 1 &  SDSSJ123902.93+654934.5 & 0 & 1 &  SDSSJ215819.27+000909.6 & 1 & 0 \\
  SDSSJ074244.88+421425.7 & 1 & 1 &  SDSSJ095043.94+391541.6 & 1 & 1 &  SDSSJ125645.47+252241.6 & 0 & 1 &  SDSSJ230202.49-000930.0 & 1 & 1 \\
  SDSSJ074301.93+410655.2 & 0 & 1 &  SDSSJ095250.46+155304.1 & 1 & 1 &  SDSSJ131630.63+612412.5 & 1 & 0 &  SDSSJ235524.29+044855.8 & 0 & 1 \\
  SDSSJ074605.00+480048.7 & 1 & 1 &  SDSSJ100811.87+162450.4 & 1 & 1 &                            &   &   \\
\hline
\end{tabular}
\end{small}
\end{table*}

We  run Kolmogorov-Smirnov  (KS) tests  to quantitatively  compare the
cumulative distributions (a $\chi^2$ test in the case of the secondary
star    spectral    types).     The   results    are    provided    in
Table\,\ref{t-probs}.  The  probabilities are in most  cases very low,
thus indicating  that the  WDMS binaries of  the four  sub-samples are
drawn from  different parent  populations.  Indeed,  there are  no two
sub-samples for which the obtained probabilities are above 15 per cent
in all  four cases (white  dwarf effective temperature,  mass, surface
gravity  and secondary  star spectral  type). The  fact that  the four
sub-samples seem to  be statistically different is  not too surprising
and clearly indicates  that the SDSS WDMS binary  catalogue is heavily
affected by selection effects of the different populations.

Legacy and BOSS WDMS binaries are most likely accidentally selected by
the target  selection algorithm of  quasars and, in principle,  we may
expect these populations to be  statistically similar.  However, as we
have already  mentioned, the  quasar selection algorithm  has modified
the  limiting  magnitude from  $i<$19.1  in  SDSS\,I/II to  $g<$22  in
SDSS\,III  \citep{rossetal12-1}.   Intrinsically faint  WDMS  binaries
containing late-type companions and/or cool/high-mass white dwarfs are
therefore  more likely  to be  observed by  the BOSS  survey than  the
Legacy  survey.  Thus,  the  BOSS WDMS  binary  population displays  a
scarcity of low-mass (\Mwd$\la0.5$\Msun) as  well as a higher fraction
of   cooler  ($\la$15000\,K)   white  dwarfs   than  the   Legacy  one
(Figure\,\ref{f-histo}).  Moreover,  the secondary star  spectral type
distribution of BOSS  WDMS binaries presents a  clear overabundance of
late-type  ($\geq$M6)  companions as  compared  to  the spectral  type
distribution of Legacy  WDMS binaries.  This result  suggests that the
observed overall  scarcity of SDSS WDMS  binaries containing late-type
companions (bottom  right panel of Figure\,\ref{f-histo})  is a simple
consequence of  selection effects  incorporated by the  SDSS selection
criteria rather than an intrinsic physical property of these binaries.

SEGUE  WDMS  binaries  were  observed thanks  to  a  dedicated  survey
performed by  us to select objects  with a strong contribution  of the
companion star  \citep{rebassa-mansergasetal12-1}, hence  they present
statistical properties that are  different from the other sub-samples.
Interestingly, white  dwarfs of both  SEGUE and SEGUE-2  WDMS binaries
seem to  be of  similar properties  in terms of  white dwarf  mass and
surface gravity (Table\,\ref{t-probs}), although the number of SEGUE-2
systems  is too  low  and these  similarities need  to  be taken  with
caution.

SEGUE-2  WDMS binaries  are  selected  as the  result  of a  selection
algorithm that  aims at obtaining  spectra of main sequence  stars and
red giants.   The colours  of WDMS  binaries are  similar to  those of
single  main sequence  stars only  when the  flux contribution  of the
white dwarf is  small, i.e.  the white dwarfs are  expected to be cool
and/or massive.   Indeed, the white  dwarf population of  SEGUE-2 WDMS
binaries is the  coolest among the four sub-samples and  the masses of
1/3 of the white  dwarfs are $\geq$0.8\,\Msun (Figure\,\ref{f-histo}).
The secondary stars that are part  of SEGUE-2 WDMS binaries seem to be
of   similar    properties   to    those   of   the    Legacy   sample
(Table\,\ref{t-probs}).

\section{Radial Velocities and identification of PCEBs}
\label{s-rvs}

We here measure the \Lines{Na}{I}{8183.27, 8194.81} absorption doublet
and  H$\alpha$ emission  radial  velocities (RV)  from  all SDSS  WDMS
binary spectra identified in this work. We analyse these RVs to detect
close binaries,  i.e.  PCEBs, in  our sample.   That is, an  object is
considered to be a PCEB if we detect significant (more than 3$\sigma$)
RV variation.  If we do not detect RV variation we consider the system
as a  likely wide  binary, as  the probability exists  for the  RVs to
sample the same  orbital phase of a  PCEB, in which case  we would not
detect RV variation  (note also that low-inclination  and long orbital
period PCEBs are more difficult to identify).

We fit  the \Lines{Na}{I}{8183.27,8194.81}  absorption doublet  with a
second order polynomial  plus a double-Gaussian line  profile of fixed
separation,  and  the H$\alpha$  emission  line  with a  second  order
polynomial  plus  a  single-Gaussian  line profile,  as  described  in
\citet{rebassa-mansergasetal08-1} and  \citet{renetal13-1}.  Each SDSS
spectrum  is  the  result  of combining  several  different  exposures
(hereafter sub-spectra), hence  we measure the RVs  from all available
sub-spectra to  thus increase the  chances of detecting  short orbital
period PCEBs. However,  we need to keep in mind  that a large fraction
of WDMS binaries found in this work are observed by the BOSS survey of
SDSS  and   that  their  spectra   are  generally  of  low   SN  ratio
(Figure\,\ref{f-sn}).  Therefore,  we expect a large  fraction of BOSS
WDMS binary sub-spectra to be of  extremely low SN ratio and hence not
suitable for measuring  accurate RVs.  Thus, we decide  to measure the
RVs from all combined spectra too. Measuring RVs from combined spectra
has the main effect of decreasing  the sensitivity to RV variations of
very  short orbital  period PCEBs  due to  averaging out  some orbital
phase information.

We  derive at  least  one  \Lines{Na}{I}{8183.27, 8194.81}  absorption
doublet  and  H$\alpha$  emission  RV   of  an  accuracy  better  than
20\,\kms\, for 536 and 291 WDMS binaries, respectively.  Since many of
these WDMS  binaries are not  new identifications, we combine  the RVs
measured here to those available  from our previous studies to analyse
the number  of close binaries,  i.e.  PCEBs,  in the sample.   We thus
count  390  and  224  WDMS   binaries  with  at  least  two  available
\Ion{Na}{I}   and/or   two   available   H$\alpha$   RV   measurements
respectively, among  which we detect  73 objects displaying  more than
3$\sigma$ \Ion{Na}{I} RV variation and 66 systems displaying more than
3$\sigma$ H$\alpha$ RV variation. If we only take into account systems
with RVs taken on different  nights\footnote{230 WDMS binaries have at
  least two  available \Ion{Na}{I} RVs  taken on different  nights, of
  which 62 display more than 3$\sigma$ RV variation.  127 objects have
  at least  two available  H$\alpha$ emission  RVs taken  on different
  nights, of  which 51 display  more than 3$\sigma$ RV  variation.} we
derive a PCEB  fraction of $\sim$27 per cent based  on our \Ion{Na}{I}
RV  measurements, or  $\sim$40  per  cent based  on  our H$\alpha$  RV
measurements.  Whilst the former value  is in agreement with the close
binary fraction found by \citet{nebotetal11-1}  of 21-24 per cent, the
latter seems  to be  rather overestimated.  This  is a  consequence of
H$\alpha$ RVs being less robust as probe for RV variations most likely
due to  magnetic activity affecting  the secondary stars, i.e.   the M
dwarfs may be  relatively rapidly rotating for  the H$\alpha$ emission
to  be  patchy  over  their  surface thus  resulting  in  apparent  RV
variations \citep{rebassa-mansergasetal08-1}.  Hence, caution needs to
be  taken  with those  PCEBs  displaying  only H$\alpha$  emission  RV
variation.  Two  obvious exceptions here  are SDSSJ093947.95+325807.3,
which is  found to  be an  eclipser (see  Section\,\ref{s-eclip}), and
SDSSJ125645.47+252241.6,  for which  we  estimate  its orbital  period
based on the  H$\alpha$ RVs (see below).  Excluding duplicate binaries
that display  both \Ion{Na}{I}  and H$\alpha$ RV  variation we  end up
with  98  unique  PCEBs,  of  which 62  are  new  identifications.  In
Table\,\ref{t-rvs} we provide the IAU names of the 62 new PCEBs.

\begin{figure}
\centering
\includegraphics[width=\columnwidth]{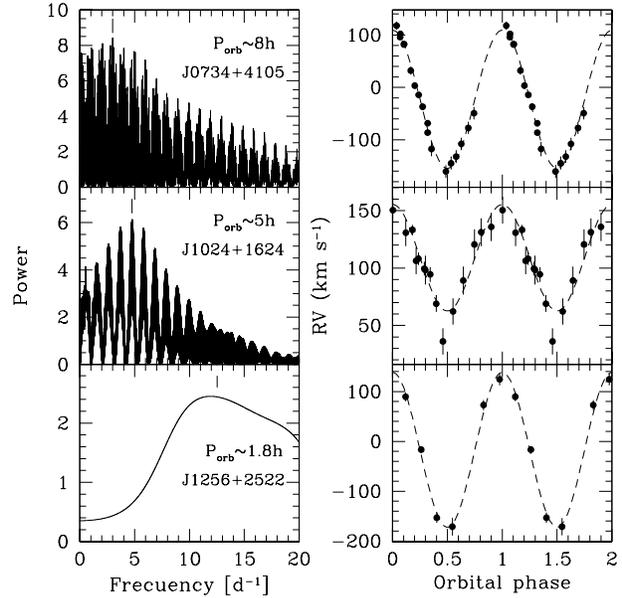}
\caption {Left: Scargle periodograms calculated from the RV variations
  measured    from    the    \Ion{Na}{I}   absorption    doublet    in
  SDSSJ073455.91+410537.4  and  SDSSJ102438.46+162458.2 and  from  the
  H$\alpha$ emission line in SDSSJ125645.47+252241.6.  The most likely
  orbital  periods  (i.e.  the  values  with  the highest  power)  are
  indicated by tick marks.  Right: the RV measurements folded over the
  adopted orbital periods of the systems.}
\label{f-periods}
\end{figure}

We finally  combine the  RV information presented  here with  those RV
values from  our previous studies  of all  SDSS PCEBs and  analyse the
\citet{scargle82-1}  periodograms  calculated  from the  RVs  of  each
system to  investigate the periodic  nature of the RV  variations.  In
three  cases   (SDSSJ073455.91+410537.4,  SDSSJ102438.46+162458.2  and
SDSSJ125645.47+252241.6) we  are able to estimate  the orbital periods
based   on   the   resulting   periodograms  (see   left   panels   of
Figure\,\ref{f-periods}). We  adopt the orbital periods  as the values
corresponding to  the highest  power in the  periodograms and  we then
carry out sine-fits of the form

\begin{equation}
\label{e-fit}
V_\mathrm{r} =
K_\mathrm{sec}\,\sin\left[\frac{2\pi(t-T_0)}{P_\mathrm{orb}}\right]
+\gamma
\end{equation}

\noindent
to the  RV data sets  of each system,  where $\gamma$ is  the systemic
velocity, $K_\mathrm{sec}$  is the  radial velocity  semi-amplitude of
the companion star,  $T_0$ is the time of inferior  conjunction of the
secondary star,  and $P_\mathrm{orb}$ is  the orbital period.   The RV
values folded  over the  estimated orbital  periods obtained  form the
sine    fits    are    provided     on    the    right    panels    of
Figure\,\ref{f-periods}. The estimated parameters resulting from these
fits are reported in Table\,~\ref{t-periodaliases}.

The RVs  measured in  this section, the  heliocentric Julian  dates of
each observation and  information on whether the systems  are PCEBs or
wide    WDMS   binaries    can   be    found   in    our   web    site
\emph{http://www.sdss-wdms.org}.

\section{Eclipsing WDMS binaries}
\label{s-eclip}

In  this final  section we  search for  eclipsing WDMS  binaries using
photometry from the  Catalina Sky Survey (CSS) and  Catalina Real Time
Transient  Survey (CRTS;  \citealt{drakeetal09-1}).   We followed  the
same    approach     detailed    in     \citet{parsonsetal13-1}    and
\citet{parsonsetal15-1}  whereby we  selected all  WDMS binaries  with
magnitudes of $r<19$ and re-reduced the raw CSS data ourselves to more
easily  identify   very  deeply  eclipsing  systems   and  remove  any
contaminated exposures.   Out of the  1,177 and 91 WDMS  binaries and
candidates  identified here,  only  314 were  bright  enough and  with
sufficient  coverage  to search  for  eclipses,  i.e.  a  fraction  of
$\sim$25 per cent. This low percentage  is mainly due to the fact that
a large  fraction of the  WDMS binaries  identified in this  work were
observed  by the  BOSS  survey of  SDSS of  limiting  magnitude $g  <$
22.0. That  is, many WDMS  binaries do not  survive the $r<19$  cut we
applied.  We identified seven eclipsing  systems, however all of these
were         previously        known:         SDSSJ011009.14+132615.66
\citep{pyrzasetal09-1},                      SDSSJ093947.92+325807.59,
SDSSJ095719.24+234240.69,                    SDSSJ141057.72-020236.48,
SDSSJ142355.06+240924.43,                     SDSSJ145634.25+161138.11
\citep{drakeatal10-1},           and          SDSSJ141536.38+011718.59
\citep{greenetal78-1}.        We       also      discovered       that
SDSSJ074244.91+421426.17  is  in  fact  a MS+MS  eclipsing  binary  as
revealed by  a deep  secondary eclipse  and thus  exclude it  from our
list. Hence,  the total number of  SDSS WDMS binaries is  3294, 646 of
which are  new.  The number of  eclipsing systems among our  sample is
consistent with previous results indicating  that $\sim$10 per cent of
PCEBs  ---  which  make  up   roughly  one  fourth  of  WDMS  binaries
\citep{schreiberetal10-1,   rebassa-mansergasetal11-1,  nebotetal11-1}
--- are eclipsing systems \citep{parsonsetal13-1}.

\begin{table}
\setlength{\tabcolsep}{1ex}
\caption{\label{t-periodaliases}     Estimated     orbital     periods
  $P_\mathrm{orb}$,  semi-amplitudes   $K_\mathrm{sec}$  and  systemic
  velocities $\gamma$. It  is worth noting the  high systemic velocity
  of SDSSJ102438.46+162458.2.}
\begin{flushleft}
\begin{center}
\begin{tabular}{cccc}\hline\hline
Object & $P_\mathrm{orb}$ & $K_\mathrm{sec}$ & $\gamma$ \\
       &   (h)          &  (\kms)         & (\kms) \\
\hline
SDSSJ073455.91+410537.4 & 8   & 132 & -23 \\
SDSSJ102438.46+162458.2 & 5   & 46  & 109 \\
SDSSJ125645.47+252241.6 & 1.8 & 155 & -17 \\
\hline
\end{tabular}
\end{center}
\end{flushleft}
\end{table}

\section{Summary and conclusions}

The spectroscopic  catalogue of SDSS  WDMS binaries now  contains 3294
systems from DR\,12.  979 are  additions from the here presented work,
646  of which  have not  been published  before.  This  is by  far the
largest,  most  complete and  most  homogeneous  catalogue of  compact
binaries  currently available.   We have  provided stellar  parameters
(white dwarf effective temperatures, surface gravities and masses, and
secondary star  spectral types) and  RVs for the here  identified WDMS
binaries.  We  have also applied  3D corrections to the  derived white
dwarf parameters  of all  SDSS WDMS  binaries.  The  stellar parameter
distributions corresponding to WDMS binaries observed by the different
surveys  of  SDSS (Legacy,  BOSS,  SEGUE,  SEGUE-2) are  statistically
different due to  the different target selection  criteria used.  This
clearly reveals  that the  SDSS WDMS binary  sample, though  being the
largest, is heavily affected by  selection effects.  In particular, we
find indications for these selection effects to be responsible for the
overall scarcity of  observed SDSS WDMS binaries  containing late type
($\geq$M6) companions.   The RVs of 98  of our here studied  SDSS WDMS
binaries  (62 of  which are  new identifications)  display significant
variation and we  flag these systems as close binaries.   For three of
them the RV data are sufficient to estimate their orbital periods.  We
have also identified seven WDMS binaries as eclipsing, although all of
them have been published before.

The SDSS WDMS  binary catalogue is a superb sample  that is being used
to tackle many different open problems in modern astrophysics, ranging
from constraining  the physical  properties of low-mass  main sequence
stars and white dwarfs to improving our understanding of close compact
binary evolution,  among others.

\section*{Acknowledgments}

We thank the  anonymous referee for his/her  comments and suggestions.
This research has been funded  by MINECO grant AYA2014-59084-P, by the
AGAUR, by  the European  Research Council  under the  European Union's
Seventh   Framework  Programme   (FP/2007-2013)/ERC  Grant   Agreement
n.320964 (WDTracer), by Milenium  Science Initiative, Chilean Ministry
of Economy, Nucleus  P10-022-F, by Fondecyt (1141269,  3140585) and by
the National Key Basic Research Program of China (2014CB845700).

Funding  for  SDSS-III  has  been  provided by  the  Alfred  P.  Sloan
Foundation,  the  Participating  Institutions,  the  National  Science
Foundation, and the  U.S. Department of Energy Office  of Science. The
SDSS-III web site is http://www.sdss3.org/.

SDSS-III is managed  by the Astrophysical Research  Consortium for the
Participating Institutions of the SDSS-III Collaboration including the
University of  Arizona, the Brazilian Participation  Group, Brookhaven
National  Laboratory,   Carnegie  Mellon  University,   University  of
Florida,  the French  Participation  Group,  the German  Participation
Group, Harvard  University, the Instituto de  Astrofisica de Canarias,
the Michigan State/Notre Dame/JINA  Participation Group, Johns Hopkins
University,   Lawrence  Berkeley   National  Laboratory,   Max  Planck
Institute for Astrophysics, Max  Planck Institute for Extraterrestrial
Physics, New Mexico State University,  New York University, Ohio State
University, Pennsylvania  State University, University  of Portsmouth,
Princeton University,  the Spanish Participation Group,  University of
Tokyo,  University  of  Utah,  Vanderbilt  University,  University  of
Virginia, University of Washington, and Yale University.

\appendix
\section{Notes on individual systems}

\begin{figure}
\centering
\includegraphics[angle=-90,width=\columnwidth]{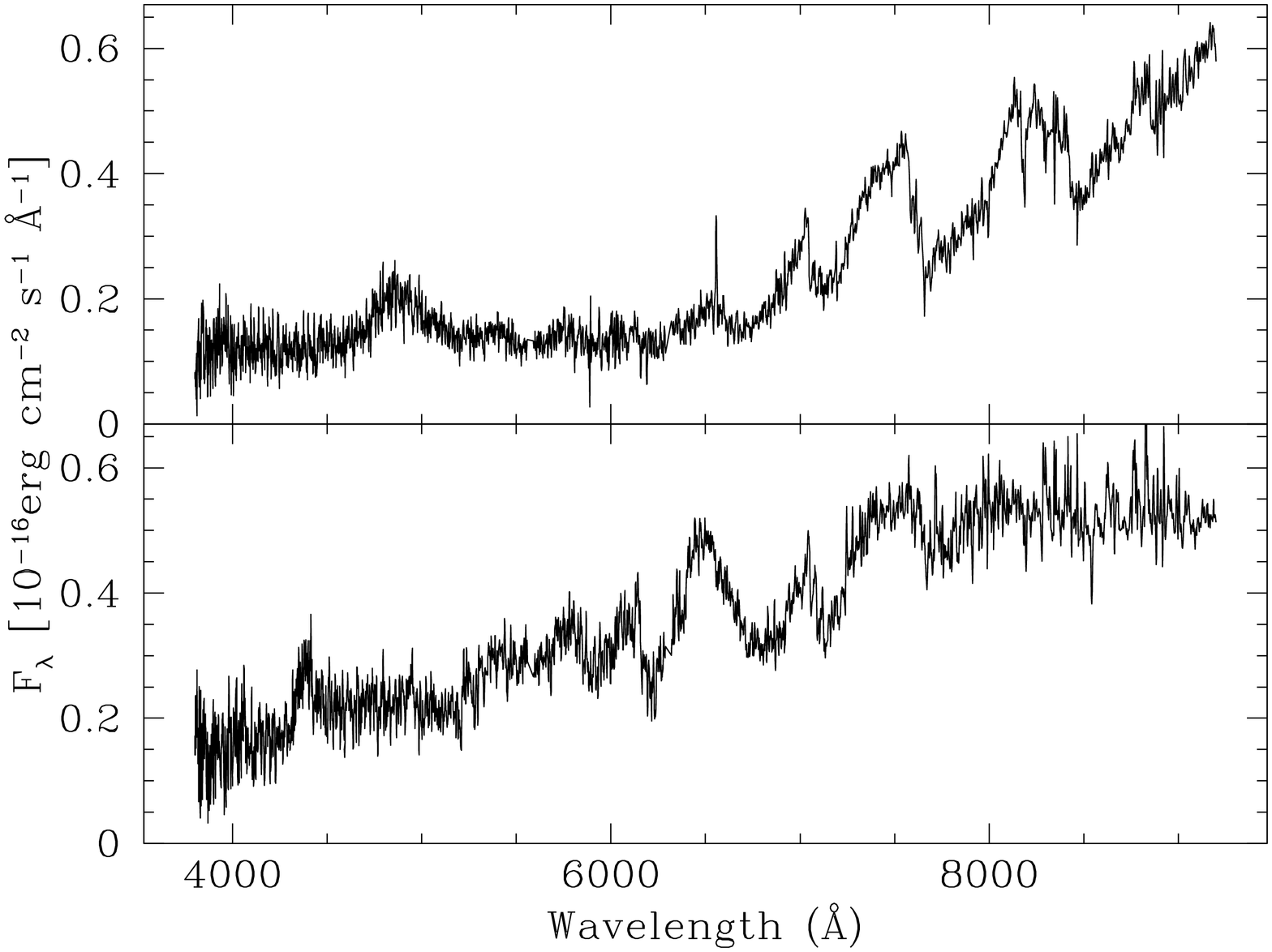}
\caption {Top  panel: SDSS spectrum of  SDSSJ013714.97+210220.0, a new
  LARP.  Bottom panel: SDSS spectrum of SDSSJ100615.27+242612.1, a new
  candidate LARP.}
\label{f-larps}
\end{figure}

\begin{figure}
\centering
\includegraphics[angle=-90,width=\columnwidth]{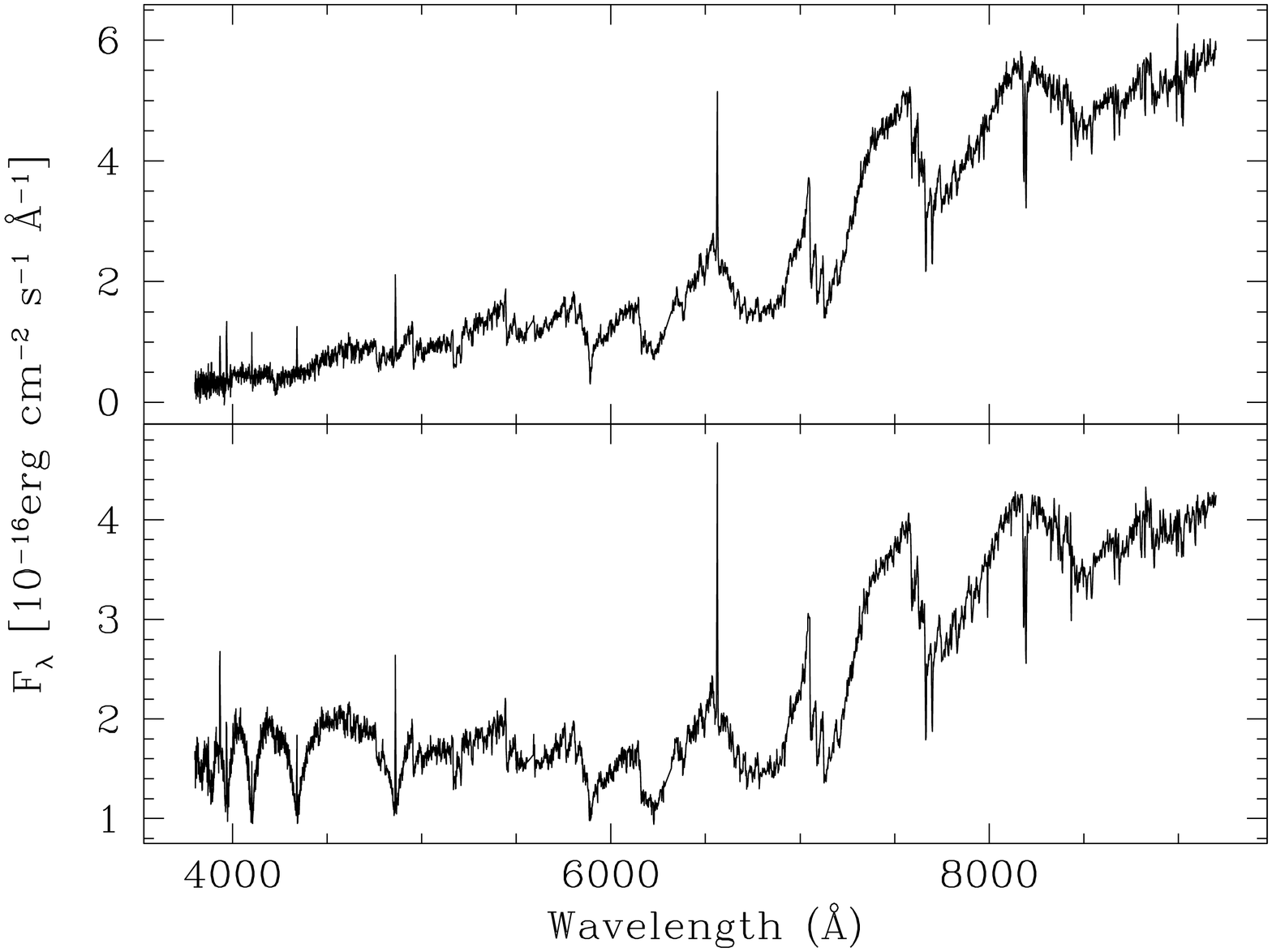}
\caption {Two SDSS spectra of SDSSJ081647.38+534017.8. The white dwarf
  is clearly visible  in the bottom panel spectrum,  but is completely
  absent in the top panel spectrum.  This may be because the top panel
  spectrum was  taken during  an eclipse. Puzzlingly,  the \Ion{Na}{I}
  RVs measured from the SDSS spectra do not show any variation.}
\label{f-0816}
\end{figure}

\begin{figure}
\centering
\includegraphics[angle=-90,width=\columnwidth]{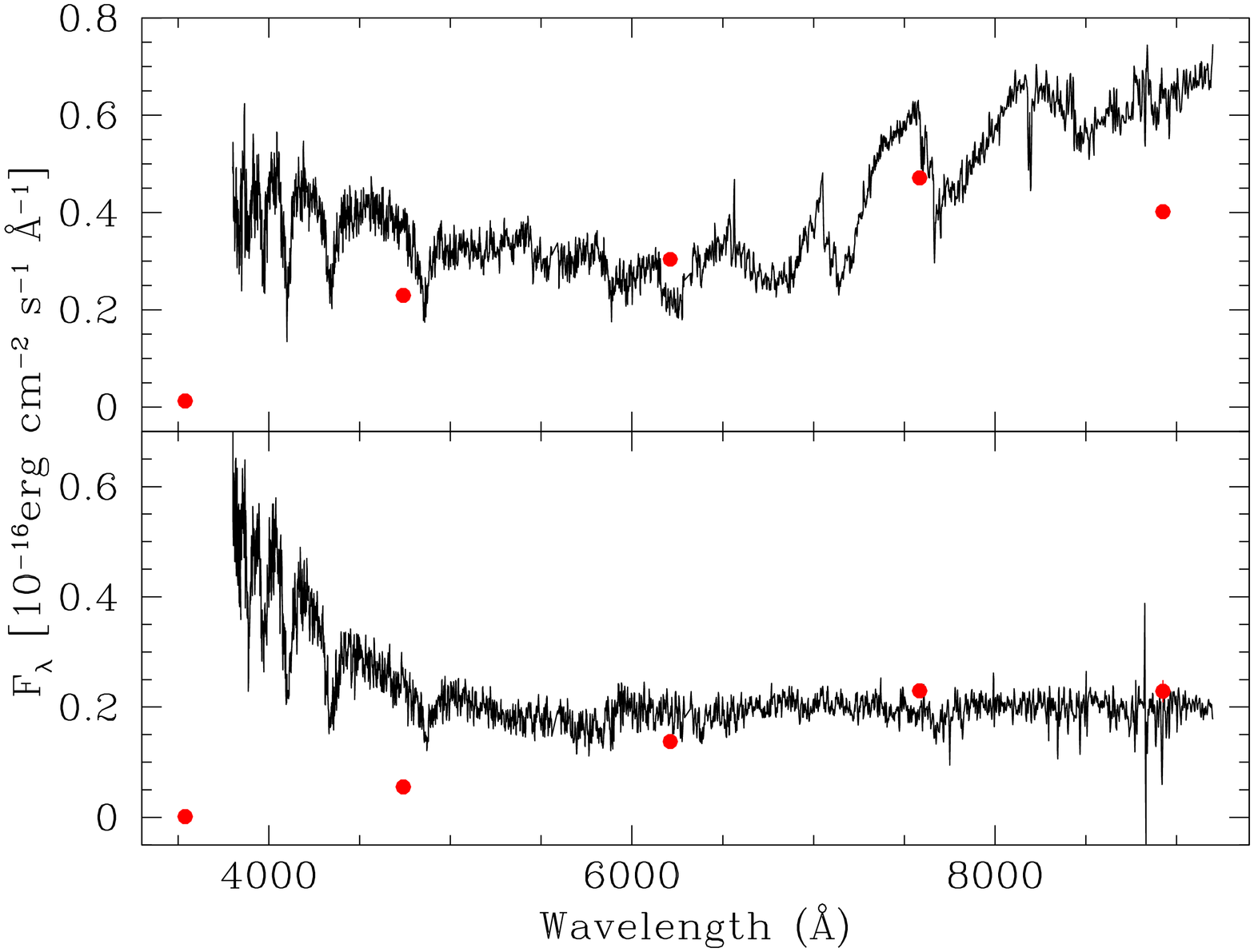}
\caption  {SDSS spectra  of  SDSSJ083938.54+395440.5  (top panel)  and
  SDSSJ161213.77+163906.2 (bottom panel).  $ugriz$ fluxes derived from
  the available SDSS photometry are shown as red solid dots.}
\label{f-uband}
\end{figure}

\emph{SDSSJ003336.49+004151.3,                SDSSJ014745.02-004911.1,
  SDSSJ024310.60+004044.4,         SDSSJ132040.27+661214.8         and
  SDSSJ132830.92+125941.4}: these  five WDMS binaries have  been found
to display  \Ion{Na}{I} RV variation  (Section\,\ref{s-rvs}). However,
they  were flagged  as  wide  WDMS binaries  in  our previous  studies
\citep{schreiberetal10-1,   rebassa-mansergasetal11-1}.     This   may
indicate that these binaries have relatively long orbital periods of a
few  hundred  of days  (as  the  spectra  analysed  in this  work  are
separated by  this amount  of time  from the  spectra analysed  in our
previous studies), or simply that our previous RV measurements sampled
always the same orbital phases.\\

\noindent \emph{SDSSJ013714.97+210220.0  and SDSSJ100615.27+242612.1}:
we classify  SDSS0137 as  a new low-accretion  rate polar  (LARP). The
SDSS spectrum displays clear  cyclotron emission at $\sim$4800\AA.  We
identify two emission bumps  at $\sim$4400\AA\, and $\sim$6500\AA\, in
the SDSS spectrum of SDSSJ1006 that may also arise as a consequence of
cyclotron emission. Hence, SDSSJ1006 may  also be a low-accretion rate
polar.   The   SDSS  spectra   of  the  two   systems  are   shown  in
Figure\,\ref{f-larps}.\\

\noindent   \emph{SDSSJ073953.89+392732.5}:  we   classify  the   main
sequence     companion    as     a     metal-poor    subdwarf.\\

\noindent    \emph{SDSSJ081647.38+534017.8,   SDSSJ083938.54+395440.5,
  SDSSJ101819.47+174702.34,  and SDSSJ161213.77+163906.2}:  these four
WDMS binaries may be eclipsing.   Two SDSS spectra of SDSSJ0816 reveal
the  presence  and absence  of  the  white dwarf,  respectively,  thus
suggesting the  second spectrum  was obtained  during an  eclipse (see
Figure\,\ref{f-0816}).  However,  it is important to  mention that the
\Ion{Na}{I} RVs measured from the SDSS  spectra do not reveal any sign
of variation.  The  SDSS spectra and $ugriz$ fluxes  derived from SDSS
photometry    of    SDSSJ0839    and   SDSSJ1612    are    shown    in
Figure\,\ref{f-uband}.  Inspection the figure clearly reveals that the
$u$-band fluxes  (also the $z$-band  flux for SDSSJ0839) do  not match
the  spectra.  We  consider  this may  be a  consequence  of the  SDSS
photometry  being   obtained  during  eclipses,  i.e.    the  $u$-band
photometry  correspond  to  that  of  the  main  sequence  companions.
Finally, we detect several faint points in the Catalina light curve of
SDSSJ1018, indicating that it may  well be eclipsing, however there is
insufficient data to confirm this or determine its period.\\

\noindent \emph{SDSSJ123931.98+210806.2}: the SDSS spectrum shows a DA
white dwarf, a late-type secondary  star and Balmer lines in emission.
Originally classified as a WDMS binary, this object was excluded from
our sample  after we realised  it is the cataclysmic  variable IR\,Com
\citep{manser+gaensicke14-1}.  The  SDSS spectrum  was taken in  a low
state.\\

\noindent \emph{SDSSJ213408.21+065057.5}:  the SDSS  spectrum displays
the typical  features of  a DO degenerate  and a  early-type secondary
star. H$\alpha$ is in emission.

\end{document}